\documentclass[twocolumn,showpacs,preprintnumbers,amsmath,amssymb]{revtex4-1}
\usepackage{chemformula} % Formula subscripts using \ch{}
\usepackage{graphicx}
\usepackage[colorlinks=true,linkcolor=blue,citecolor=blue]{hyperref}
\usepackage{color}
\usepackage{braket}
\usepackage{physics}

\renewcommand{\vec}[1]{\boldsymbol{#1}}
\usepackage{ulem}

\begin{document}

\title{Unveiling the Enhancement of Spontaneous Emission at Exceptional Points}

\author{L. Ferrier$^1$}\email{lydie.ferrier@insa-lyon.fr}
\author{P. Bouteyre$^1$}\thanks{L. Ferrier and P. Bouteyre contributed equally to this work as first authors.}
\email{paul.bouteyre@ec-lyon.fr}
\author{A. Pick$^2$}
\author{S. Cueff$^1$}
\author{N.H.M Dang$^1$}
\author{C. Diederichs$^3$}
\author{A. Belarouci$^1$}
\author{T. Benyattou$^1$}
\author{J.X. Zhao$^4$}
\author{R. Su$^4$}
\author{J. Xing$^5$}
\author{Q. Xiong$^{6,7}$}
\author{H. S. Nguyen$^{1,8}$}
\email{hai-son.nguyen@ec-lyon.fr}

\affiliation{$^1$Univ Lyon, Ecole Centrale de Lyon, CNRS, INSA Lyon, Universit\'e Claude Bernard Lyon 1, CPE Lyon, CNRS, INL, UMR5270, 69130 Ecully, France} 
\affiliation{$^2$Applied Physics Department , Hebrew University}
\affiliation{$^3$Laboratoire de Physique de l’\'Ecole normale sup\'erieure, ENS, Universit\'e PSL, CNRS, Sorbonne Universit\'e, Universit\'e de Paris, F-75005 Paris, France}
\affiliation{$^4$Division of Physics and Applied Physics, School of Physical and Mathematical Sciences, Nanyang Technological University,Singapore 637371, Singapore}
\affiliation{$^5$College of Chemistry and Molecular Engineering, Qingdao University of Science and Technology, Qingdao 266042 China}
\affiliation{$^6$State Key Laboratory of Low-Dimensional Quantum Physics and Department of Physics, Tsinghua University, Beijing 100084, P.R. China.}
\affiliation{$^7$Beijing Academy of Quantum Information Sciences, Beijing 100193, P.R. China.}
\affiliation{$^8$Institut Universitaire de France (IUF)}

\date{\today}

\begin{abstract}

Exceptional points (EPs), singularities of non-Hermitian physics where complex spectral resonances degenerate, are one of the most exotic features of nonequilibrium open systems with unique properties. For instance, the emission rate of quantum emitters placed near resonators with EPs is enhanced (compared to the free-space emission rate) by a factor that scales quadratically with the resonance quality factor. Here, we verify the theory of spontaneous emission at EPs by measuring photoluminescence from photonic-crystal slabs that are embedded with a high-quantum-yield active material. While our experimental results verify the theoretically predicted enhancement, it also highlights the practical limitations on the enhancement due to material loss. Our designed structures can be used in applications that require enhanced and controlled emission, such as quantum sensing and imaging.  % < 125 words
\end{abstract}

\pacs{}

\maketitle

Exploring and taming open, non-conservative systems  has  always been a major challenge in physics. This relates to a plethora of problems from classical to quantum phenomena: the damping of a pendulum's swing by sliding friction, coherent light escaped  from the cavity of a diode laser, harnessing thermal radiation for radiative cooling, and decoherence mechanisms in quantum systems. The past few years have witnessed the triumph of non-Hermtiticy as the modern approach to describe non-conservative mechanisms in a broad range of open systems~\cite{Rotter2015,El-Ganainy2018,ElGanainy2019}. These systems, theoretically described by non-Hermitian hamiltonians , would exhibit peculiar features with no Hermitian counterparts. One may cite the non-Hermitian extension of topological matter~\cite{RevModPhys.93.015005}, and the formation of bound states in the continuum resulted from destructive interference of losses~\cite{Hsu2016}. Exceptional points (EPs) are prototypical examples of a unique degeneracy that can occur in non-Hermitian systems \cite{Berry1998,Heiss1999,Regensburger2012,Ge2012} in which at least two eigenvectors and associate complex eigenvalues simultaneously coalesce. Fundamentally, EPs represent singularities of non-Hermitian topology \cite{Shen2018,Kawabata2019,Sone2020}. For instance, in the case of isolated EPs, two eigenstates can be swapped when adiabatically encircling an EP in the parameter space \cite{Zhou2018,Gao2015,Doppler2016,Liu2020}, a direct consequence of the isolated EPs' half topological charges. Due to their topological nature, many other intriguing phenomena were discovered in systems with EPs such as unidirectional transmission or reflection \cite{Regensburger2012,Lin2011,Peng2014}, loss-induced transparency \cite{Guo2009}, topological chirality \cite{Doppler2016,Xu2016}, chirality-reversal radiation \cite{Chen2020}. For devices applications, novel concepts for making sensors with higher sensibility \cite{Wiersig2014,Hodaei2017,Chen2016,Chen2017EP,Park2020,Dong2019} and lasers with intriguing properties \cite{Miao2016,Gao2017,Gu2016,Hodaei2014,Feng2014,Peng2016,Liertzer2012,Peng2014b,Brandstetter2014} using EP properties have been suggested and implemented.

\begin{figure*}[!ht] 
\centering
   \includegraphics[width=0.9\linewidth]{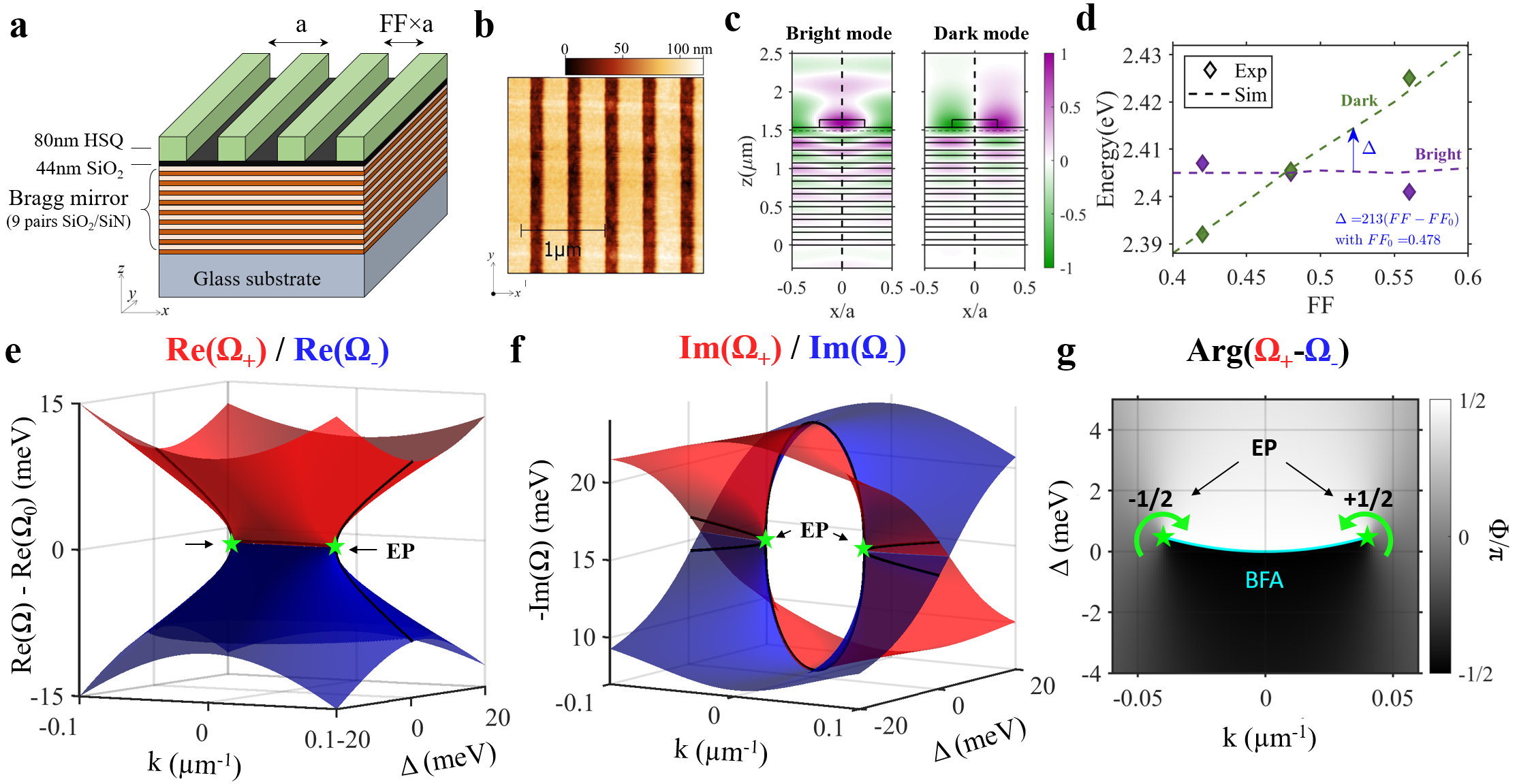}
  \caption{\textbf{Sample and analytical model.} \textbf{a}, Sketch of the passive sample composed of a hydrogen silsesquioxane (HSQ) grating on top of a Bragg mirror (9 $SiO_2$/SiN pairs) with a 44 nm spacing layer of SiO2. The grating period $a$ is 500 nm, with a spacing of $a \times FF$, with $0<FF<1$ the filling factor. \textbf{b}, Atomic Force Microscopy (AFM) image of the passive sample. \textbf{c}, Electric field distribution simulation of the bright and dark modes at wavevector $k=0$. \textbf{d} Energy of the dark and bright modes as a function of the filling factor (FF). The star symbols are experimental data and the lines are results obtained from RCWA simulations. \textbf{e} and \textbf{f}, Real and imaginary parts of the system eigenvalues as a function of the wavevector $k_x$ in the direction of the grating, and the energy gap $\Delta$ between the two modes. The two EPs, i.e. simultaneous degeneration of the eigenvalues' real and imaginary parts, are indicated with green stars. \textbf{g}, Argument of the complex gap, $\Phi=Arg(\Omega_+-\Omega_-)$, revealing the  half topological charges of EPs. The two yellow points indicate the two isolated EPs and the blue line the Bulk Fermi Arc (BFA).}
 \label{Figure1}
\end{figure*}

Recent theoretical work unravelled the mystery regarding the apparent divergence of the emission enhancement at EPs (i.e., the so-called  Peterman factor) and predicted unique spectral features with substantial but finite enhancement at EPs \cite{Lin2016,Pick2017}. However, experimental verification of these results was missing until very recently \cite{Wang2020}. In ref\,\cite{Wang2020}, the authors show that for the application of sensing, the enhancement of the Petermann factor near the EP is accompanied by a commensurate increase in the noise signal;  Therefore, the signal-to-noise ratio is not dramatically improved near the EP and that limits the applicability of the EP effect in gyro-based sensing applications. While LDOS  (Local Density of State) enhancement near EPs offers limited improvement in sensing capabilities\cite{Wang2020}, the implication of EPs for enhanced emission is much more promising    since the enhancement of spontaneous emission near EPs in actively pumped structures is theoretically unbounded\,\cite{Pick2017}. Here, we report on the first experimental demonstration of spontaneous emission enhancement at EPs. In particular, we design an experimental platform to demonstrate and analyze the enhancement at EPs taking full account of realistic constraints. The EPs are directly observed from angle-resolved reflectivity measurements, and their LDOS enhancement is revealed via photoluminescence signal when a high-quantum-yield active material is implemented into the system. A finite LDOS enhancement factor was measured and is in perfect agreement with the analytical value predicted by recently proposed theory on LDOS at EPs\,\cite{Pick2017} in the framework of an analytical non-Hermitian model.  This result is an essential step towards using EPs in applications as engineering LDOS is at the heart of most of light-matter interaction mechanisms such as accelerating and directing spontaneous emission, tailoring light-harvesting efficiency, enhancing photonic nonlinearity and molding photonic transport.

%Here, we  investigate experimentally the LDOS enhancement at isolated photonic EPs. The EPs are directly observed from angle-resolved reflectivity measurements, and their LDOS enhancement is revealed via photoluminescence signal when a high quantum yield active material is implemented into the system. A finite LDOS enhancement factor  was measured verifying  the  theoretically  predicted  enhancement. Furthermore, the experimental enhancement factor is in perfect agreement with the analytical value predicted by recently proposed theory on LDOS at EPs \cite{Pick2017} in the frame work of an analytical non-Hermitian model. 

\begin{figure*}[!ht]
\centering
\includegraphics[width=0.9\linewidth]{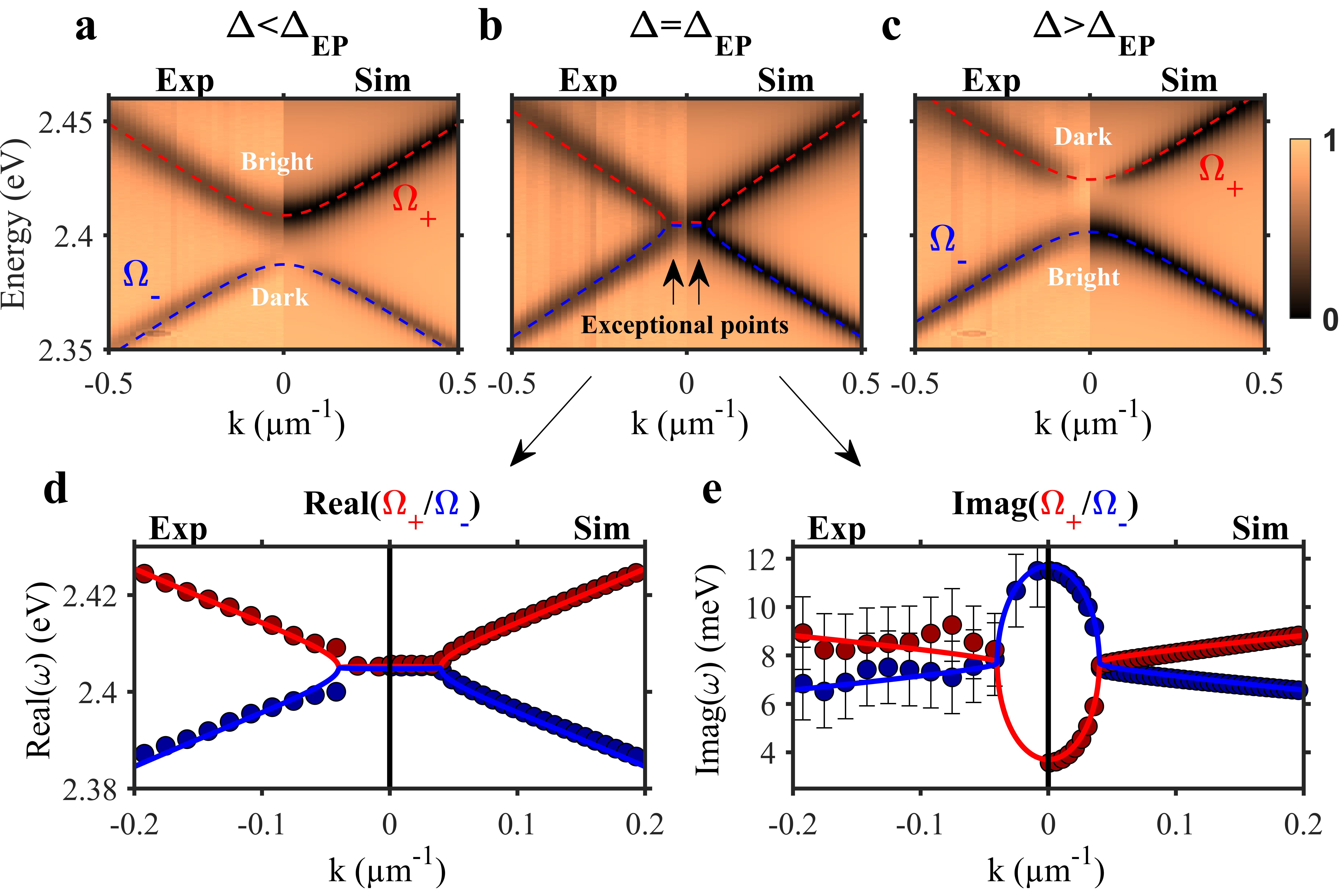}
  \caption{\textbf{Experimental investigation on two isolated EPs.} \textbf{a} to \textbf{c}, Angle-resolved reflectivity  measurements (left panels) and corresponding numerical simulations (right panels) of the sample  for $\Delta=-21$ meV (\textbf{a}), $\Delta=0.48$ meV (\textbf{b}) and $\Delta=24$ meV (\textbf{c} ). \textbf{d} and \textbf{e}, Real and imaginary parts of the experimental (left, dots) and simulated (right, dots) eigenvalues of the structure with EPs in \textbf{b} compared with the analytical model (lines) given by \eqref{eq1}. At $\Delta=\Delta_{EP}$ and for $k<\abs{k_{EP}}$, the real part of the eigenvalues is almost degenerated and it is only possible to extract the bright branch from the experimental data. The parameters used for the analytical models are $\Delta= 0.48 \: meV$, $E_0=2.405 \: eV$, $v= 100 \: meV .\mu m^{-1}$, $\delta= -300 \: meV .\mu m^{-2}$, $\gamma_b=7.5 \: meV$, and $\gamma_{nr}=3.8 \: meV$. The exceptional points are located at $\pm 0.04 \: \mu m^{-1}$.}
  \label{Figure2}
\end{figure*}

To engineer isolated EPs, we employ subwavelength unidimensional (1D) photonic lattices exhibiting lateral mirror symmetry $-x\rightarrow x$ (see Figures \ref{Figure1}\textbf{a} and \textbf{b}) and study the band structures in the vicinity of the first  bandgap at $\Gamma$ point. The two eigenmodes of this gap are Transverse Electric (TE) modes of opposite parities, and are accordingly denoted in the following as dark (antisymmetric) and bright (symmetric) modes (see Figure \ref{Figure1}\textbf{c}). The dark mode cannot couple to the radiative continuum and corresponds to a symmetry protected bound state in the continuum with zero radiative losses \cite{Lee2019,Lu20}. Consequently, by playing with a set of two or more uncorrelated parameters, it is possible to make the dark and bright modes coalesce into EPs \cite{Zhen2015,Lee2019,Lu20}. Using the dark and bright states as basis, the eigenmodes in the vicinity of $\Gamma$ can be described by a non-Hermitian $k\cdot p$ Hamiltonian (more information in the Supplementary Material \,\cite{supp}): 
\begin{equation}
	\label{eq1}
	H(k,\Delta)=E_0+
	\left(\begin{matrix} 
	\frac{\Delta+ \delta.k^2}{2}  & v.k\\
    v.k & - \frac{\Delta+ \delta.k^2}{2}
 \end{matrix} \right)
 +i \left(\begin{matrix} 
	\gamma_{nr} + \gamma_b  & 0\\
    0 & \gamma_{nr} 
 \end{matrix} \right),
\end{equation}

In the Hermitian term of \eqref{eq1},  $\Delta$ is the energy gap between the dark and bright modes, $E_0$ the mid-gap energy, $\delta$ and $v$ the two coefficients of $k \cdot p$ perturbation theory when second order is included. In the non-Hermitian term of \eqref{eq1}, $\gamma_b$ is the radiative loss of the bright mode and $\gamma_{nr}$ the nonradiative loss of both modes. Interestingly, together with the wavevector $k$,  the energy gap $\Delta$ can be implemented as a synthetic dimension, to form a two-dimensional parameter space $\vec{q}=(k,\Delta)$. These two parameters are effectively independent of each other as the wavevector $k$ is related to the angle of far-field radiation, whereas the energy gap $\Delta$ is dictated by the grating filling factor $FF$ (see \ref{Figure1}\textbf{d}).

The mapping of the real and imaginary parts of the eigenvalues of \eqref{eq1} are plotted in Figures \ref{Figure1}\textbf{e} and \ref{Figure1}\textbf{f}, respectively. One can observe that they are simultaneously degenerated at two EPs of coordinates: $k_{EP}=\pm\frac{\gamma_b}{2v_g}$, and $\Delta_{EP} = -\delta.\abs{k_{EP}}$. The isolation and topological nature of these EPs are revealed from the texture of the phase $\Phi (k,\Delta)$ which is defined by the argument of the two eigenvalues complex difference, $\Phi=arg(\Omega_+-\Omega_-)$ \cite{Shen2018,Kawabata2019}. As shown in Figure \ref{Figure1}\textbf{g}, one can observe two isolated EPs, both possessing half topological charges, in the 2D synthetic space.  Indeed, encircling each EPs accumulates a vortex phase that is equal to $\pm \pi$ (see Figure \ref{Figure1}\textbf{g}), with corresponding winding numbers $\frac{1}{2\pi} \oint_\mathcal{C} d\vec{q} \nabla_{\vec{q}} \Phi=\pm\frac{1}{2}$~\cite{Shen2018,Kawabata2019}. Finally, these two EPs are connected by a bulk Fermi arc, given by  $\Delta=-\delta k^2$, along which the real part of eigenvalues are degenerated~\cite{Zhou2018}.\\ 

Reflectivity experiments are performed to evidence the two isolated EPs predicted by the analytical model. Figures \ref{Figure2} \textbf{a}-\textbf{c} present the experimental angle-resolved reflectivity maps (left panels) and the numerically simulated ones (right panels), for three different values of $\Delta$: (\textbf{a}) $\Delta<\Delta_{EP}$, (\textbf{b}) $\Delta=\Delta_{EP}$, and (\textbf{c}) $\Delta>\Delta_{EP}$. In Figures \ref{Figure2} \textbf{a} and \textbf{c}, one can easily identify the dark mode for which the radiative resonance vanishes at $k=0$ due to its antisymmetric parity.  These figures also evidence the inversion of the two bands when the difference $\Delta-\Delta_{EP}$ switchs sign. Importantly, for $\Delta= \Delta_{EP}$  in Figure \ref{Figure2} \textbf{b}, the two bands coalesce at two EPs located at $k_{EP}=\pm 0.04 \: \mu m^{-1}$. To further confirm the EPs formation, the real and imaginary parts of the eigenvalues are respectively retrieved from the spectral position of the resonance dips and their linewidths.  These experimental values are depicted in Figures \ref{Figure2} \textbf{d} and \textbf{e}, which exhibit a very good agreement with the numerical simulations results and are nicely reproduced by the analytical model.  

\begin{figure*}[ht!]
\centering
  \includegraphics[width=0.9\linewidth]{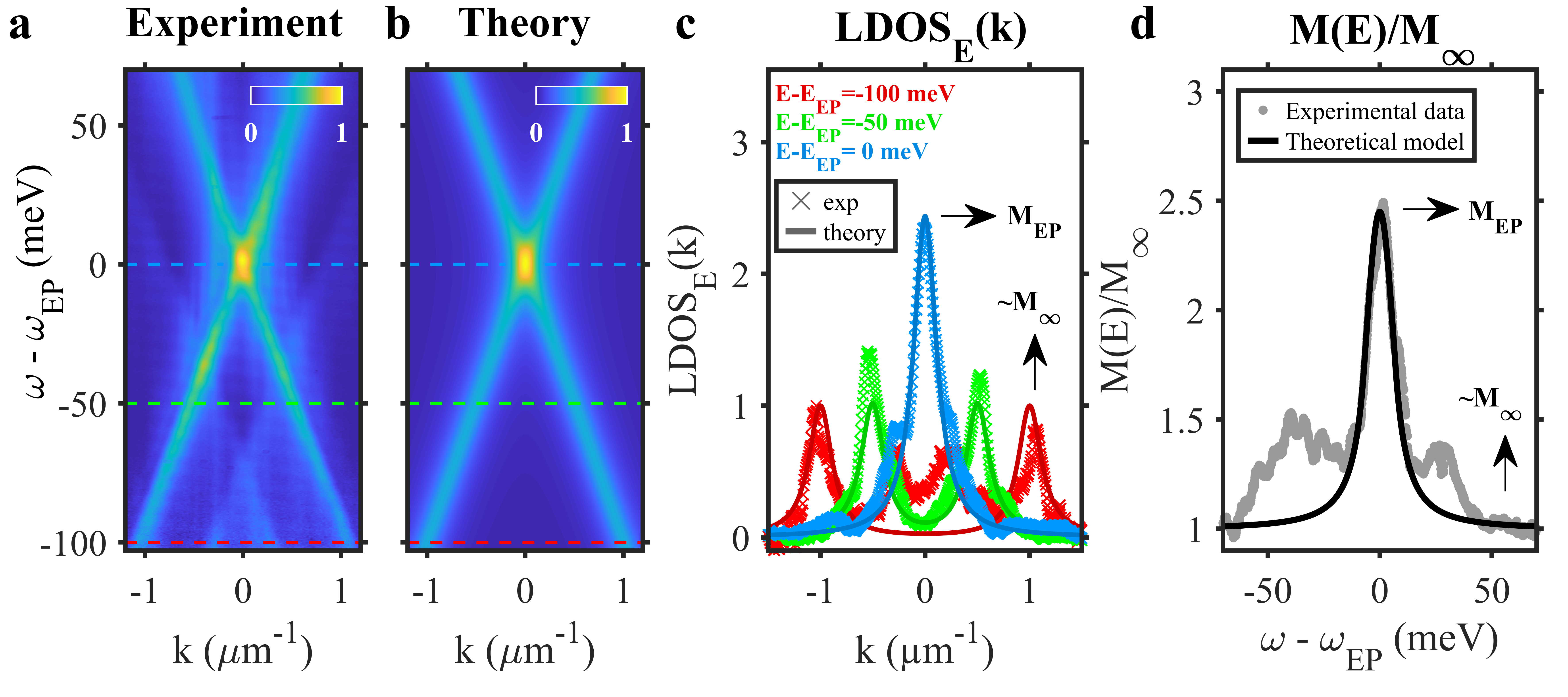}
  \caption{\textbf{LDOS enhancement at the EPs}. \textbf{a}, Experimental LDOS map extracted from the photoluminescence measurements of the active sample (sample in figure \ref{Figure1}a with an extra 15 nm-!thick \ch{CsPbBr_3} layer.). \textbf{b} Theoretical LDOS map obtained by using the model on the LDOS at EPs in \cite{Pick2017} and our analytical model. The parameters used for the theoretical LDOS are $\Delta= 0 \: meV$, $E_0=2.3878 \: eV$, $v= 100 \: meV .\mu m^{-1}$, $\delta= 0 \: meV .\mu m^{-2}$, $\gamma_b=5 \: meV$, and $\gamma_{nr}=11 \: meV$. \textbf{c}, Three horizontal isofrequency cross-section, at $\omega-\omega_{EP}$=-100, -50, and 0 meV, of the experimental (crosses) and theoretical (lines) LDOS of \textbf{a} and \textbf{b}. $M_{EP}$ and $M_\infty$, peak values at the energy of respectively the EPs and the non-degenerate resonances (i.e. away from the EPs energy) are indicated. \textbf{d}, Experimental (gray dots) and theoretical (black line) LDOS peak values at given energies, $M(E)=\max_k(LDOS_E(k))$, normalized by the LDOS peak value at the non-degenerate resonances away from the exceptional points, $M_{\infty}$.   } 
 \label{Figure4}
\end{figure*}

To probe the LDOS at the isolated EPs, a 15 nm-thick layer of \ch{CsPbBr_3} perovskite colloidal nanocrystals is deposited on the sample. The choice of this active material is based on two advantages: first, they can sustain near-unity photoluminescence quantum efficency at the EPs wavelength \cite{DiStasio2017,GUALDRONREYES2021499}, thus the LDOS is directly proportional to the emission intensity; and second, they can be easily implemented into the passive structure by spin-coating with good uniformity. 

Figure \ref{Figure4}\textbf{a} presents the experimental mapping of the LDOS in the vicinity of the EPs, extracted from the angle-resolved photoluminescence measurements. Theoretical predictions, obtained by implementing the Hamiltonian \eqref{eq1} to the LDOS model from Ref. \cite{Pick2017}, is depicted in Figure \ref{Figure4}\textbf{b} and reproduce remarkably well the experimental measurements. One can observe from both the experimental and theoretical maps that the LDOS resonate with the photonic modes. Most importantly, the LDOS signal is maximized around the EPs (i.e. at $E=E_{EP}$ and $k=0$), revealing an enhancement of the LDOS at EPs. Note that the two EPs can no longer be distinguished from one another as the non-radiative losses, $\gamma_{nr}$, broaden the two EPs LDOS peaks.  

Further insights of the LDOS enhancement are gained by examining its distribution in momentum space at given energy $LDOS_E(k)$. Such distribution is simply obtained from isofrequency cross-section of the LDOS map.  As an illustration, Figure \ref{Figure4}\textbf{c} shows three cross-sections of the experimental and theoretical LDOS of figure \ref{Figure4}\textbf{a} and \textbf{b}, corresponding to $E-E_{EP}$=-100 meV, -50 meV, and 0 meV. As for the LDOS maps, we have again a very good agreement between experiment and theory.  From these momentum-resolved distributions, the LDOS peak $M(E)= \max_k\left[LDOS_E(k)\right]$ is retrieved.  Figure \ref{Figure4}\textbf{d} compares the experimental (gray dots) and theoretical (black line) spectrally-resolved LDOS peaks. The profile of the experimental $M(E)$ in the vicinity of the EPs energy is nicely matched to the theoretical $M(E)$ in terms of linewidth and amplitude. For energies away from the EPs energy, a deviation occurs between experiment and theory. This is explained by the coupling of the nanocrystals emission to additional band-folded Bragg modes \,\cite{supp} which are not considered in our effective theory. Finally, particular attention is paid to the LDOS peaks at the EPs, $M_{EP}$, and away from the EPs, $M_\infty$, both indicated in Figures \ref{Figure4}\textbf{c} and \textbf{d}. The experimental LDOS enhancement corresponds then to the ratio between $M_{EP}$ and $M_\infty$ giving a value of 2.56. This experimental finding answers to the fundamental question of the LDOS enhancement at EPs as the enhancement remains finite despite the non-orthogonality of the eigenvectors as predicted in recent theories \cite{Pick2017}. Furthermore, the experimental results shown in figure \ref{Figure4} are well explained by the theory on LDOS at EPs \cite{Pick2017}.

In our system, the modal degeneracy at the EP produces an enhancement factor of 2.56. This implies that the intensity at the EP is 2.56 times stronger than that of a  single non-degenerate resonance (which is enhanced by the traditional Purcell factor). The excess emission comes from the degeneracy and the  non-orthogonality of the modes. Since the enhancement at an ordinary degeneracy is bounded by 2, the fact that our enhancement factor exceeding 2 proves the presence of an  EP \footnote{The emission lineshape at an  EP is a squared Lorentzian while that of an ordinary degeneracy is a Lorentzian multiplied by the degree of degeneracy; in this case 2}. This factor of 2.56 can be improved by using  high-order passive EPs\,\cite{Lin2016}. Alternatively, by adding gain, this value can increase  also at second-order EPs\,\cite{Pick2017}. Although increasing the gain would inevitably increase $\gamma_{nr}$, recent work shows that it is possible to achieve high gain with low loss  by utilizing  hybrid light-matter polaritonic modes that arise from the strong-coupling regime between excitons of quantum wells and photons in photonic crystal\,\cite{Ardizzone2022}.

In conclusion, the recent theoretical predictions on the LDOS enhancement at EPs \cite{Pick2017} have been confirmed experimentally while taking full account of realistic constraints from a photonic-crystal slab platform. In our experiment, a finite enhancement factor of 2.56 has been measured and is in good agreement with the one given by our analytical theory. Our results open the way to LDOS engineering in non-Hermitian photonics for novel optoelectronics devices such as lasing operating at EPs when important gain medium is introduced, or nonlinear optics harnessing LDOS enhancement\cite{nonlinearEPlasing}. Finally, while this work only studies the radiation of an ensemble of quantum dots at EPs, the effect of EPs on spontaneous emission of single quantum emitters ~\cite{Chen2020} is a salient perspective to explore new regime of cavity-quantum electrodynamics for novel single photon sources. For example, by performing temporal dynamic experiments on single emitters, our platform could be used to demonstrate the recent prediction from ref\,\cite{Khanbekyan2020} of  increased  lifetime of quantum excitations near EPs. \\

\textit{Acknowledgement:}
This work was partly funded by the French National Research Agency (ANR) under the project POPEYE (ANR-17-CE24-0020) and the IDEXLYON from Université de Lyon, Scientific Breakthrough project TORE within the Programme Investissements d’Avenir (ANR-19-IDEX-0005). It is also supported by the Auvergne-Rh\^{o}ne-Alpes region in the framework of PAI2020 and the Vingroup Innovation
Foundation (VINIF) annual research grant program under Project Code VINIF.2021.DA00169. Q.X. gratefully acknowledges the funding support from National Natural Science Foundation of China (No. 12020101003) and Tsinghua University start-up grant.  The authors thank Xavier Letartre, Pierre Viktorovitch and Xuan Dung Nguyen for fruitful discussions.

\onecolumngrid

\begin{center}
	\textbf{\large --- SUPPLEMENTAL MATERIAL ---}
\end{center}

\newcommand{\figref}[1]{Fig.~\ref{fig:#1}}
\newcommand{\secref}[1]{Sec.~\ref{#1}}
\renewcommand{\eqref}[1]{Eq.~(\ref{eq:#1})}
\newcommand{\eqsref}[2]{(\ref{eq:#1},\ref{eq:#2})}
\renewcommand{\vec}[1]{\mathbf{#1}}
\newcommand{\vecg}[1]{\boldsymbol{#1}}
\newcommand{\mat}[1]{\mathbb{#1}}
\renewcommand{\appendixname}{APPENDIX}
\newcommand{\appendixsec}[2]{\section{\uppercase{#1}}\label{#2}}
\newcommand{\appendixsubsec}[2]{\subsection{\uppercase{#1}}\label{#2}}
\renewcommand{\mat}[1]{\overline{\overline{#1}}}

\renewcommand{\theequation}{S\arabic{equation}}
\renewcommand{\thefigure}{S\arabic{figure}}

\section{Theory of LDOS at exceptional points}

In this section,  we present a brief recap of the theory for spontaneous emission near EPs and, then, proceed to derive a simplified expression that captures the key features of our experimental results [\eqref{GreenEPexample}].  Most generally, the rate of spontaneous emission is determined by the number of electromagnetic modes that an emitter can emit into, given by  the local density of states (LDOS). The latter is proportional to the imaginary part of the Green's function. Hence, our goal in this section is to obtain a simple expression for the Green's function. 

We will be using a non-Hermitian formulation of the problem, which takes into account radiation loss by  imposing outgoing boundary conditions to solve Maxwell's equations.  Describing  Maxwell's equations formally   by an  operator $H(\vec{r})$, the  Green's function is the system's response to a point source excitation, given by the following relation:
\begin{align}
[H(\vec{r})-\omega\mathbb{I}]G(\vec{r},\vec{r}',\omega) = -\delta(\vec{r},\vec{r}').
\end{align}
For convenience of discussion, let us transform  the partial differential equation  into matrix notation (by choosing an appropriate basis). We denote matrices by overlines and vectors by bold letters. Electromagnetic resonant modes  are obtained by solving the eigenvalue problem
\begin{align}
\mat{H}\vec{x}_n = \omega_n\vec{x}_n,
\end{align}
where outgoing wave solutions are imposed  in the construction of  $\mat{H}$.
Assume for simplicity that $\mat{H}$ is symmetric.
When all its eigenvalues are semi-simple (i.e., the spectrum does not contain EPs), the modal expansion of $\mat{G}$ is \cite{Arfken2006}
\begin{align}
\mat{G}(\omega) = \sum_n \frac{1}
{\omega-\omega_n}
\frac{\vec{x}_n \vec{x}_n^T}{\vec{x}_n^T \vec{x}_n},
\label{eq:FullExpansion}
\end{align}
where the superscript   $T$ denotes (unconjugated) transposition.
Assume that two of the  resonances ($\omega_\pm$) are nearly degenerate and, in addition, are   spectrally separated from all the other resonances ($\omega_n$). 
(Formally, we require  that $\left|\mathrm{Re}(\omega_n-\omega_\pm)\right|\gg
\max\{\left|\mathrm{Im}(\omega_n)\right|,\left|\mathrm{Im}(\omega_\pm)\right|\}$.) For frequencies near the degeneracy  $\omega\approx\mathrm{Re}(\omega_\pm)$, one can approximate $\mat{G}$ by keeping two poles in \eqref{FullExpansion}:
\begin{align}
\mat{G}(\omega) \approx 
\frac{1}{\omega-\omega_+}
\frac{\vec{x}_+ \vec{x}_+^T}{\vec{x}_+^T \vec{x}_+}
+
\frac{1}{\omega-\omega_-}
\frac{\vec{x}_- \vec{x}_-^T}{\vec{x}_-^T \vec{x}_-}.
\label{eq:GreenNearEP}
\end{align}
Let the Hamiltonian depend linearly on a scalar parameter $p$
\begin{align}
\mat{H}(p)=\mat{H}_0 + p\mat{H}_1,
\label{eq:TaylorEpxpansion}
\end{align}
and assume that $\omega_\pm(p)$ and $\vec{x}_\pm(p)$ coalesce at $p=0$, [thus forming a second order EP at $\omega_0\equiv\omega(p=0)$]. 
Since two eigenvectors merge into a single vector at the EP2, the Hilbert space cannot be spanned by the  eigenvectors of $\mat{H}$, but one can form a complete basis by introducing an additional vector, $\vec{j}_0$,  which satisfies
\begin{align}
\mat{H}_0\vec{x}_0 &= \omega_0\vec{x}_0\nonumber\\
\mat{H}_0\vec{j}_0 &= \omega_0\vec{j}_0 + \vec{x}_0.
\label{eq:ChainRelations}
\end{align}
where $\vec{x}_0\equiv\vec{x}(p=0)$ and $\mat{H}_0\equiv\mat{H}(p=0)$.  The second equation immediately implies that $\vec{x}_0^T\vec{x}_0=0$ (which can be seen by multiplying both sides from the left by $\vec{x}_0^T$).
To uniquely determine $\vec{x}_0$ and $\vec{j}_0$, we need two addition normalization conditions and we require $\vec{x}_0^T\vec{j}_0 = 1$ and $\vec{j}_0^T\vec{j}_0  = 0$.  
As shown in~\cite{Hernandez2003}, the modal expansion at an EP2 is
\begin{align}
\mat{G}_0(\omega) \approx 
\frac{1}{(\omega-\omega_0)^2}
\frac{\vec{x}_0 \vec{x}_0^T}{\vec{x}_0^T \vec{j}_0}
+
\frac{1}{\omega-\omega_0}
\frac{\vec{x}_0 \vec{j}_0^T+\vec{j}_0 \vec{x}_0^T}{\vec{x}_0^T \vec{j}_0},
\label{eq:degenerateG}.
\end{align}

We proceed by obtaining a simplified expression for the case of study in this present paper. In Sec. I of this supplementary, we show that  near the EP, spontaneous emission can be understood in terms of analyzing a $2\times2$ matrix [Eq. (S1)].  By subtracting  a constant   from its diagonal, the matrix can be written in the form 
\begin{align}
\mat{H} = 
\left( \begin{array}{cc}
\Omega + i\alpha & \kappa \\
\kappa & \Omega + i\alpha - i\gamma \end{array} \right),
\end{align}
The operator $\mat{H}$ has an EP2 at $\kappa = \gamma/2$, where the degenerate eigenvalue is
$\omega_0 = \Omega - i\gamma_\mathrm{EP}$, with $\gamma_\mathrm{EP}\equiv  \tfrac{\gamma}{2} - \alpha$. The vectors $\vec{X}_0 = \{i,1\}$ and $\vec{J}_0 = \{2/\gamma,0\}$ satisfy the chain relations \eqref{ChainRelations}. 
In order to implement the LDOS formula \eqref{degenerateG}, we introduce  new chain  vectors 
$\vec{x}_0 = \theta\, \vec{X}_0$ and $\vec{j}_0 = \theta\, (\vec{J}_0 + \beta\vec{X}_0)$, which  satisfy \eqref{ChainRelations} as well as   the normalization conditions $\vec{x}_0^T\vec{j}_0 = 1$ and $\vec{j}_0^T\vec{j}_0 = 0$, where $\theta = \sqrt{\gamma/(2i)}$ and $\beta = i/\gamma$. Substituting $\vec{x}_0$ and $\vec{j}_0$  into \eqref{degenerateG}, we find  that the first diagonal entry of the Green's function at the EP is
 \begin{align}
\mat{G}_0(\omega)[1,1] = 
\frac{i\gamma/2}{\left[\omega-\Omega + i\gamma_\mathrm{EP} \right]^2}+
\frac{1}
{\omega-\Omega + i\gamma_\mathrm{EP}}
\label{eq:GreenEPexample}
\end{align}
Let us consider the case where the uncoupled basis states (i.e., the eigenvectors $\vec{x}_{1,2}$ of $\mat{H}$ at $\kappa=0$) are spatially localized  at different areas in space. (A conceptually simple example is resonant modes of two uncoupled   resonators, but such points can be found in the photonic-crystal example as well \cite{Pick2016}.) In this regime, one can show that the LDOS at  spatial locations where the first mode dominates is approximately   $\mathrm{LDOS}_0(\omega) \approx -\mathrm{Im}\{\mat{G}_0(\omega)[1,1]\}$~\cite{Taflove2013}. On resonance (i.e., at $\omega = \Omega$), the LDOS peak at the EP is
 \begin{align}
M_0\equiv\max_\omega{\{\mathrm{LDOS}_0(\omega)\}}= 
\frac{\gamma/2}{\gamma_\mathrm{EP}^2}+
\frac{1}{\gamma_\mathrm{EP}}.
\label{eq:epLDOSpeak}
\end{align}
In the large coupling (nondegenerate)  limit (i.e., $\kappa\gg\gamma$),  one can use the non-degenerate modal expansion formula of the Green's function \eqref{GreenNearEP} to show that the LDOS is
 \begin{align}
 \mathrm{LDOS}_\kappa(\omega)= 
\frac{\gamma_\mathrm{EP}/2}
{\left(\omega-\Omega-\kappa\right)^2+\gamma_\mathrm{EP}^2}
+
\frac{\gamma_\mathrm{EP}/2}
{\left(\omega-\Omega+\kappa\right)^2+\gamma_\mathrm{EP}^2}
\end{align}
In this limit, the LDOS peaks at the non-degenerate resonant frequencies $\omega = \Omega\pm\kappa$ are
 \begin{align}
M_\kappa\equiv\max_\omega{\{\mathrm{LDOS}_\kappa(\omega)\}}= 
\frac{1}{2\gamma_\mathrm{EP}}.
\end{align}

\subsection*{Four-fold enhancement, $Q^2$ scaling and squared Lorentzian lineshape}
For passive systems  (with $\alpha = 0$), the resonance width at the EP is $\gamma_\mathrm{EP} = \gamma/2$ and the LDOS peak at the EP is thus four times larger than the LDOS peaks of the uncoupled resonators (i.e.,  $M_0/M_\kappa = 4$).  In the high-gain limit, where $\gamma_\mathrm{EP}\ll\gamma$, the quadratic term in \eqref{epLDOSpeak} dominates, resulting in a $Q^2$ scaling of the LDOS, in contrast to the usual $Q$ scaling of Purcell enhancement for  non-degenerate resonances. 
[Here $Q\equiv -\mathrm{Re}[\omega_n]/(2\mathrm{Im}[\omega_n])$ is the quality factor, which is a dimensionless measure of the cavity lifetime.  Note that the EP resonance width is $\gamma_\mathrm{EP}$, not to be confused with the passive width $\gamma$ in the numerator in \eqref{epLDOSpeak}.] Last, note that for high $Q$ resonances ($\mathrm{Im}[\omega_n]\ll\mathrm{Re}[\omega_n]$), \eqref{GreenEPexample} implies that the LDOS lineshape is a squared Lorentzian, in contrast to the standard Lorentzian lineshape near non-degenerate resonances. 

\section{Effective theory of 1D non-Hermitian photonic lattice} 

\subsection*{System description and the $k\cdot p$ Hamiltonian}

We consider a generic 1D photonic lattice of period $a$, corrugated along $x$ direction and invariant by translation along $y$ direction. The photonic modes confined in the lattice can leak to the radiative continuum through $z$ direction.  These are Bloch resonances, and the gap openings at $k=0$ is due to the diffractive coupling between counter propagating guided modes which are brought to the $\Gamma$ points thanks to the band-folding mechanism. Such coupling results in two eigenmodes of opposite parity with respect to the mirror symmetry $\sigma_x$. The symmetric one is leaky and can couple to the radiative continuum while the anti-symmetric one cannot couple to the radiative continuum, corresponding to a symmetry protected Bound State in the Continuum. They are noted ``bright state'' and ``dark state'' respectively in the following.\\

From the ``bright state'' and ``dark state'' at $\Gamma$ point, the $k\cdot p$ Hamiltonian is given by:

\begin{equation}
	H_{k\cdot p}(k)=\omega_0+
	\left(\begin{matrix} 
	\frac{\Delta+ \delta.k^2}{2}  & v.k\\
    v.k & - \frac{\Delta+ \delta.k^2}{2}
 \end{matrix} \right)
 +i \left(\begin{matrix} 
	\gamma_{nr} + \gamma_b  & 0\\
    0 & \gamma_{nr} 
 \end{matrix} \right)
 	\label{eq:H_kp}
\end{equation}
where $\Delta$ is the detunning between the dark and bright modes; $\delta$ and $v$ are the two coefficients of $k \cdot p$ perturbation theory when second order is included; $\gamma_{nr}$ is the nonradiative loss and $\gamma_b$ is the  radiative loss of the bright mode.

\vspace*{-10pt}

\subsection*{Eigenvalues and Exceptional Points configuration}

The complex eigenvalues of the Hamiltonian \ref{eq:H_kp} are given by:
\begin{equation}
\omega_{\pm}+i\gamma_{\pm} = \omega_0 + i\gamma_{nr} + i\frac{\gamma_b}{2} \pm \frac{1}{2}\sqrt{(\Delta + \delta.k^2 + i\gamma_b)^2 + 4v^2k^2}.
\end{equation}
The corresponding complex band gap is $\Omega_g=\sqrt{(\Delta + \delta.k^2 + i\gamma_b)^2 + 4v^2.k^2}$. The condition to obtain Exception Points (i.e. complex degeneracy) is:
\begin{equation}\label{eq:EPcondition}
(\Delta + \delta.k^2 + i\gamma_b)^2 + 4v^2.k^2 = 0
\end{equation} 
The imaginary part of Eq. \ref{eq:EPcondition} imposes $k^2=-\frac{\Delta}{\delta}$. Implementing this relation to the real part of Eq. \ref{eq:EPcondition}, we obtain another relation for $k^2$ at Exeptional Points:
$k^2=\frac{\gamma_b^2}{4v^2}$. Finally, the two conditions to achieve Exeptional Points are:
\begin{subequations}
\begin{align}
    k&=\pm\frac{\gamma_b}{2v},\\
    \Delta&=-\delta\frac{\gamma_b^2}{4v^2}.
\end{align}
\end{subequations}

\vspace*{-10pt}

\subsection*{Synthetic dimension and Topological charge}

We extend the 1D system into 2D by using $\Delta$ as a synthetic dimension. The parameter space is now given by the couple $(k,\Delta)$. With such two dimensional system, from the previous section, we know that there are two exceptional points which are pinned at:
\begin{subequations}
\begin{align}
    EP_1&=\left(-\frac{\gamma_b}{2v},-\delta\frac{\gamma_b^2}{4v^2}\right),\\
    EP_2&=\left(\frac{\gamma_b}{2v},-\delta\frac{\gamma_b^2}{4v^2}\right).
\end{align}
\end{subequations}
We now define the phase texture $\phi(k,\Delta)$ in this two dimensional space as the argument of the complex band gap:
\begin{equation}
    \phi(k,\Delta)= arg(\Omega_g).
\end{equation}
As shown in Fig.~\ref{fig:topoCharge}, the two Exceptional Points are isolated in the 2D space and corresponding to two topological half-charges: encircling each Exceptional Points provide a vortex phase amounts to $\pm \pi$.

\subsection*{Bulk Fermi Arc}

The two Exceptional Points are connected by a Bulk Fermi Arc which is characterized by the degeneracy of the real part of the eigenvalues. The equation of the Bulk Fermi Arc is obtained by imposing that the complex bandgap is purly imaginary: $\Omega_g = iB$ with $B$ a real number. Indeed, this constraint leads to:  
\begin{equation}
(\Delta + \delta.k^2 + i\gamma_b)^2 + 4v^2.k^2 = -B^2.
\end{equation}
Thus the equation of Bulk Fermi Arc corresponds to the configuration in which the imaginary part of the left term vanishes:
\begin{equation}\label{eq:bulkFermi}
\Delta=-\delta k^2
\end{equation}\\

\begin{figure}[!hb]
	\begin{center}
	\includegraphics[width=1 \textwidth]{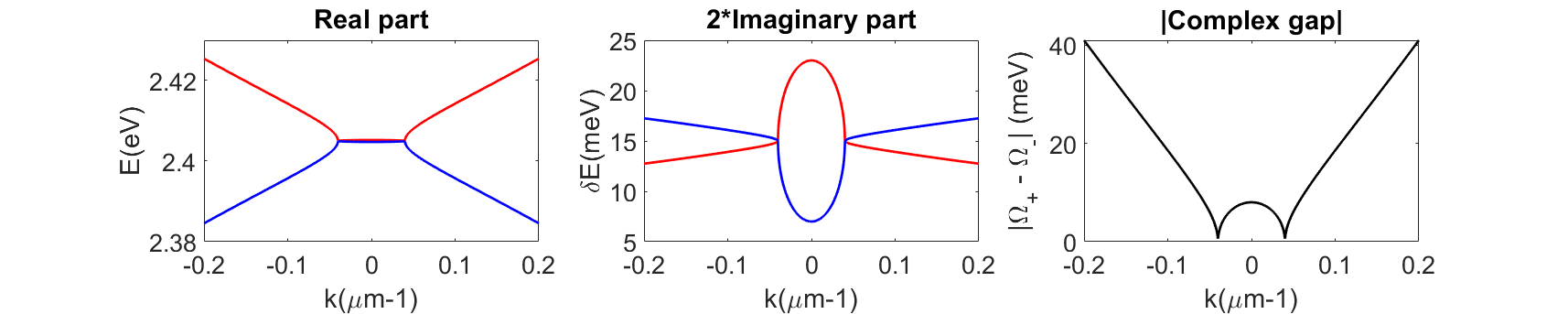}	
\caption{Exceptional Point example with  $\omega_0=2.405\,eV$, $\Delta=0.48\,meV$, $\delta=-0.3\,eV\mu m^{-2}$, $v=0.1\,eV\mu m^{-1}$, $\gamma_b=8\,meV$,  $\gamma_{nr}=3.5\,meV$.}
	\label{fig:example}
\end{center}
\end{figure}

\begin{figure}[!hb]
	\begin{center}
	\includegraphics[width=0.5 \textwidth]{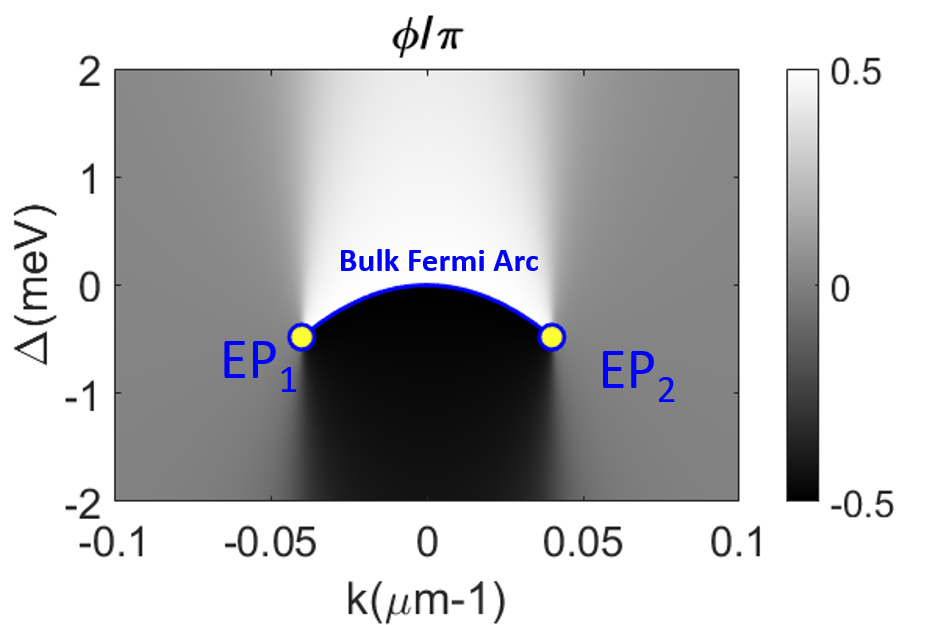}	
\caption{\label{fig:topoCharge}Phase $\phi=arg(\Omega_g)$ texture in the two dimensional parameter space $(k,\Delta)$. The other parameters are fixed: $\omega_0=2.405\,eV$, $\delta=-0.3\,eV\mu m^{-2}$, $v=0.1\,eV\mu m^{-1}$, $\gamma_b=8\,meV$,  $\gamma_{nr}=3.5\,meV$. The two Exceptional Points are half-vortices of opposite vortex number, connected by a Bulk Fermi Arc.}
\end{center}
\end{figure}

\newpage
\newpage

\begin{figure*}
\centering
\includegraphics[width=0.9\linewidth]{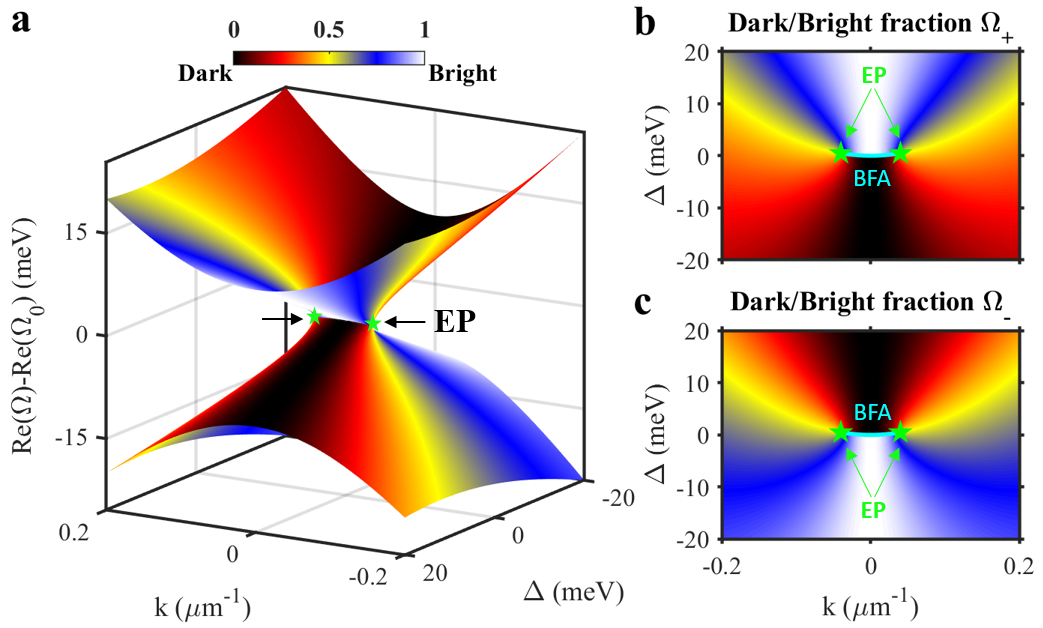}
  \caption{\textbf{Dark and Bright weights of the two system eigenvalues (a), $\Omega_{+}$ (b), and $\Omega_{-}$ (c)}}
  \label{FigureB}
\end{figure*}

\newpage 
\cleardoublepage

\section{Sample Fabrication} 

\subsection*{Bragg mirror fabrication}

The distributed Bragg reflector (DBR) was entirely deposited in a single process step in an Oxford PlasmaLab radio frequency plasma enhanced chemical vapor deposition (RF-PECVD) system. Silicon-rich silicon nitride (Si-rich SiN) and silicon dioxide (SiO2) can be deposited in the same deposition chamber by alternating the injected precursor species. Si-rich SiN with a refractive index of 2.07 at 550nm, measured by spectroscopic ellispometry, was used as a high index material, while SiO2 (n=1.46) provided a low refractive index. Both materials were deposited by exciting the plasma with a frequency of 13.54 MHz (process temperature 300 °C). The Si-rich SiN layers were deposited with a power of 70 W, a pressure of 1.5 torr, a SiH4/NH3/N2 mixed flow rate of 200/15/600 sccm, and the SiO2 layers were deposited with a power of 20 W, a pressure of 1 torr, a SiH4/N2O mixed flow rate of 100/420 sccm. The 8 SiO2 / Si-rich SiN bilayers were deposited on a c-Si substrate cleaned according to RCA standard process followed by the deposition of a 140 nm thick SiO2 spacer to target a stop-band wavelength centred around 550nm.\\

\subsection*{Grating fabrication}

The 1D photonic lattice patterns were written into an 80 nm-thick hydrogen silsesquioxane (HSQ) negative photoresist layer using e-beam lithography with a 30 keV beam. After the e-beam writing, the exposed features transform into SiO$_2$. The resulting samples are subsequently developed in a solution of 25wt\% tetramethylammonium hydroxide (TMAH) in water, maintained at a temperature of 80°C, to remove the unexposed HSQ parts.
\\

\subsection*{\ch{CsPbBr_3} QD synthesis}

CsPbBr3 QDs were synthesized according to previous report \cite{Protesescu2015}. Preparation of Cs-oleate: Cs2CO3 (0.2 g) was loaded into 100 mL 3-neck flask containing 10 mL octadecene (ODE) and 1 mL oleic acid (OA). The solution was dried for 1h at 120 $^\circ$C, and then heated under N2 to 150 $^\circ$C until all Cs2CO3 reacted with OA. The Cs-oleate/ODE precursor has to be preheated to 100 $^\circ$C before using. 
Synthesis of CsPbBr3 QDs: 5 mL ODE and 0.069 g PbBr2 were added into 25 mL 3-neck flask and dried under vacuum for 1h at 120 $^\circ$C. 0.5 mL dried oleylamine (OLA) and 0.5 mL dried OA were injected at 120 $^\circ$C under N2. After complete solubilisation of PbBr2, the solution was heated to 160 $^\circ$C and the Cs-oleate solution (0.4 mL) was quickly injected. After 5 seconds, the reaction mixture was cooled by the water bath. The product was washed with methylacetate/octane for 2 times, and finally dispersed in octane.\\

\begin{figure*}[!ht]
\centering
\includegraphics[width=0.66\linewidth]{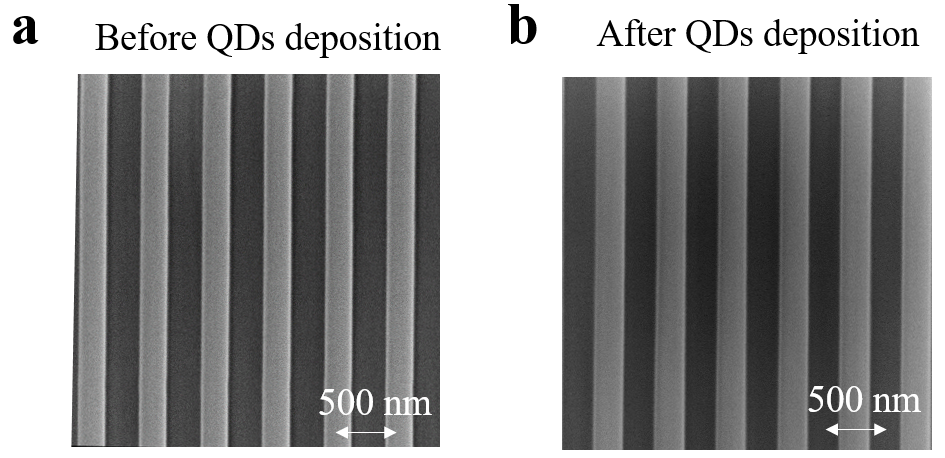}
  \caption{\textbf{Scanning Electron Microscopy (SEM) images of the second sample before and after the perovskite deposition} }
  \label{FigureF}
\end{figure*}

\newpage

\section{Thickness and refractive indices measurements}

The deposition thickness and refractive index of SiN and SiO2 via PECVD, and HSQ resist via spin-coating, have been calibrated carefully by ellipsometry measurements. The calibrations have been done by depositing single layers onto silicon substrate. The thickness of HSQ pattern in the final sample has been also measured by AFM.

\section{Numerical simulations}

The energy-momentum dispersion in figure 2 a,b, and c were simulated numerically using Rigorous Coupled-Wave Analysis (RCWA) \cite{Liu2012,Moharam1986,Alonso-Alvarez2018}. The Electric field distribution in figure 1 c and figure S9 b, and the eigenvalues real and imaginary parts in figure 2 d, and e were simulated numerically using Finite Element Method (FEM) in Comsol software. For both cases, the architecture of the simulated sample is the same as the experimental sample depicted in figure 1 a. \\

\section{Experimental setup} 
\begin{figure*}[!ht]
\centering
\includegraphics[width=\linewidth]{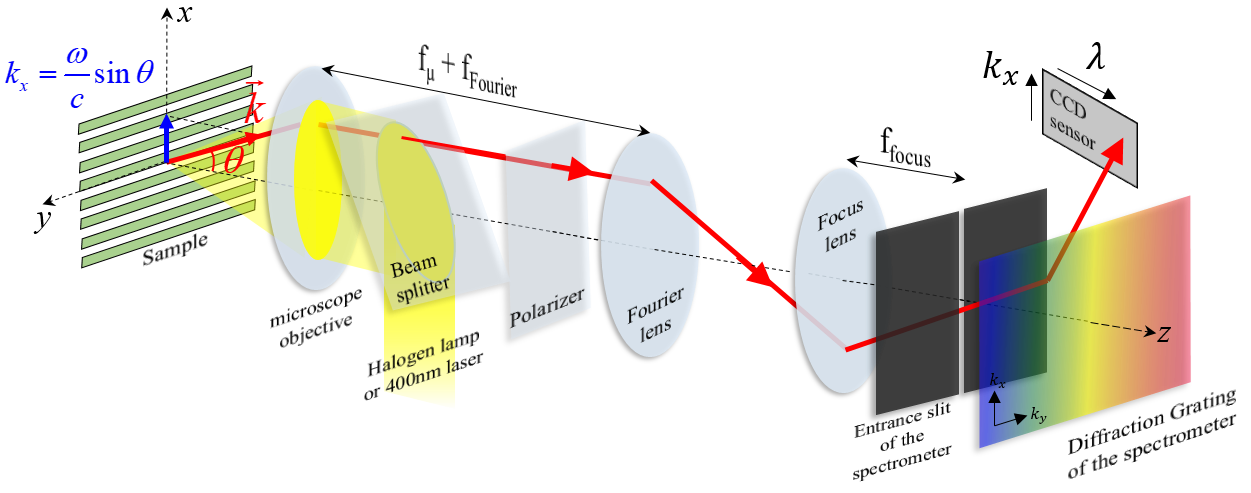}
  \caption{\textbf{Experimental setup}}
  \label{Figure_exp_set_up}
\end{figure*}

 The experimental energy-momentum dispersions are measured by a home-made setup of angle-resolved reflectivity and photoluminescence (see Fig. \ref{Figure_exp_set_up}). The Fourier plane of the grating in the back focal plane of the microscope objective (0.42 NA) is projected to the spectrometer using the "Fourier" and "Focus" lenses. The spectrometer slit selects the $k_x$-direction information, and the spectrometer diffraction grating diffracts the light in the y-direction, resulting in a ($k_x$,$\lambda$) dispersion in the spectrometer CCD sensor. For the reflectivity measurements in figure 2, the sample is shone by a Halogen lamp. For the photoluminescence data in figure 3, the sample is excited with a  400 nm pulsed laser with a repetition rate of 1kHz, resulting from frequency-doubling by a nonlinear BBO crystal from an amplifier laser source (Libra, Coherent company, center wavelength: 800 nm).

\newpage 

\section{Band folding of the Bloch Surface Waves (BSW) and Bragg modes}

\begin{figure*}[!ht]
\centering
\includegraphics[width=0.88\linewidth]{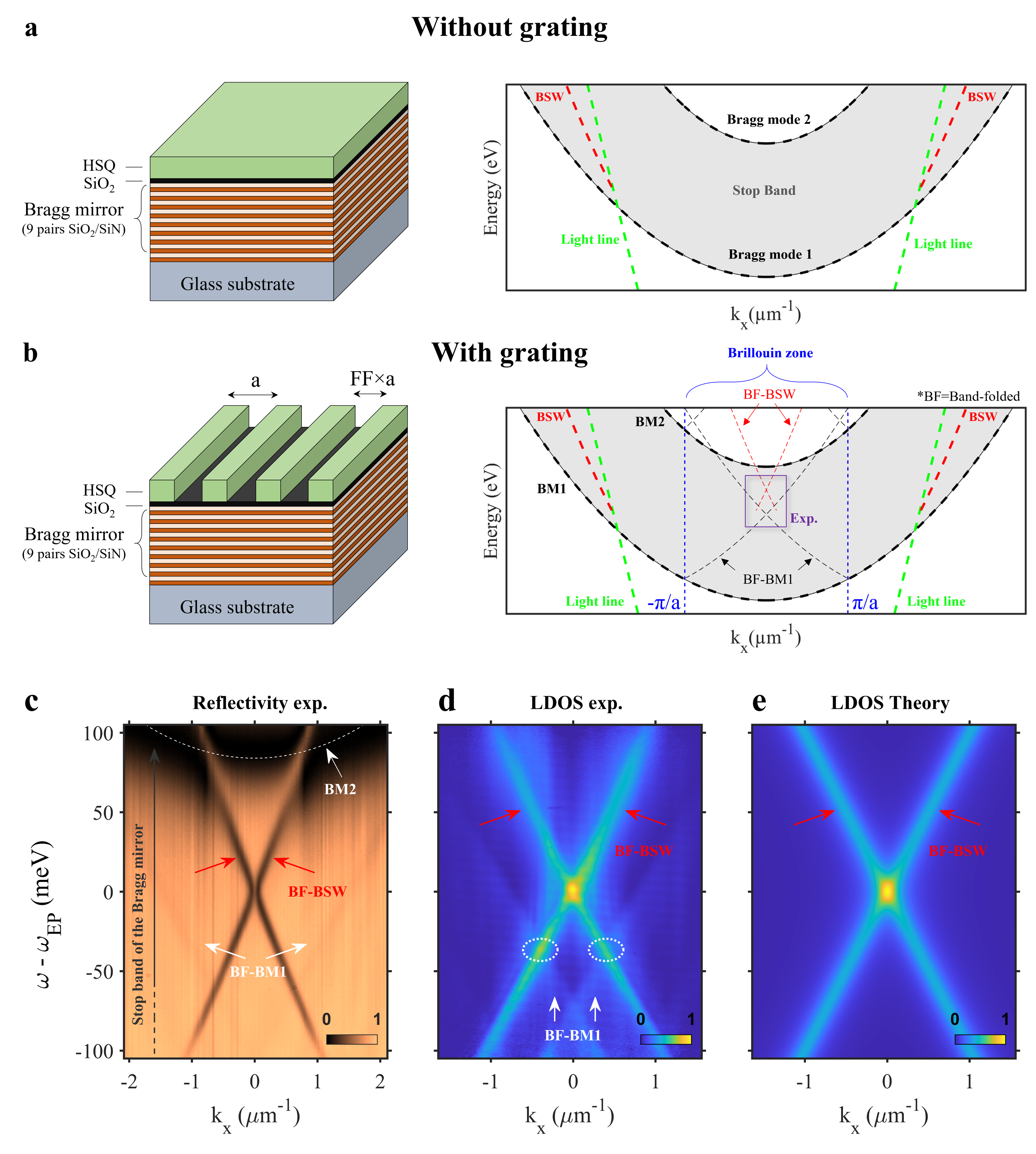}
  \caption{\textbf{Band folding of the Bloch Surface Waves (BSW) and Bragg modes}
  \textbf{a} Without any grating, the BSW modes (red bold dashed lines), propagating at the interface between the SiO2 layer and the Bragg mirror, lie under the light cone (green dashed lines) and within the Bragg mirror stop-band (gray shaded area) where the Bragg mirror is highly reflective. \textbf{b} With the hydrogen silsesquioxane (HSQ) grating, the BSW modes (red bold dashed lines) and the Bragg modes (black bold dashed lines) are band-folded at $k=\pm\pi/a$, with $a$ the grating period. The band-folded BSW modes (BF-BSW, red thin dashed lines) and the band-folded Bragg modes (BF-BM, black thin dashed lines) are located above the light cone and can therefore couple with external electromagnetic modes. The Bright and Dark modes mentioned in the article correspond to the diffractive coupling of the band-folded BSW modes. The purple box corresponds to the region of energy and wavevector of the measurements done in this study. \textbf{c} to \textbf{e} Band folding of the BSW and Bragg modes observed from \textbf{c} reflectivity measurements, \textbf{d} LDOS measurements, and \textbf{e} LDOS simulations. The white dashed circles in \textbf{e} indicate the increase of the measured LDOS compared to the predicted one due to the presence of emission coupled to the BFBM.}
  \label{FigureA}
\end{figure*}

\newpage

\section{Extraction of the eigenvalues real and imaginary parts}

\begin{figure*}[!ht]
\centering
\includegraphics[width=0.8\linewidth]{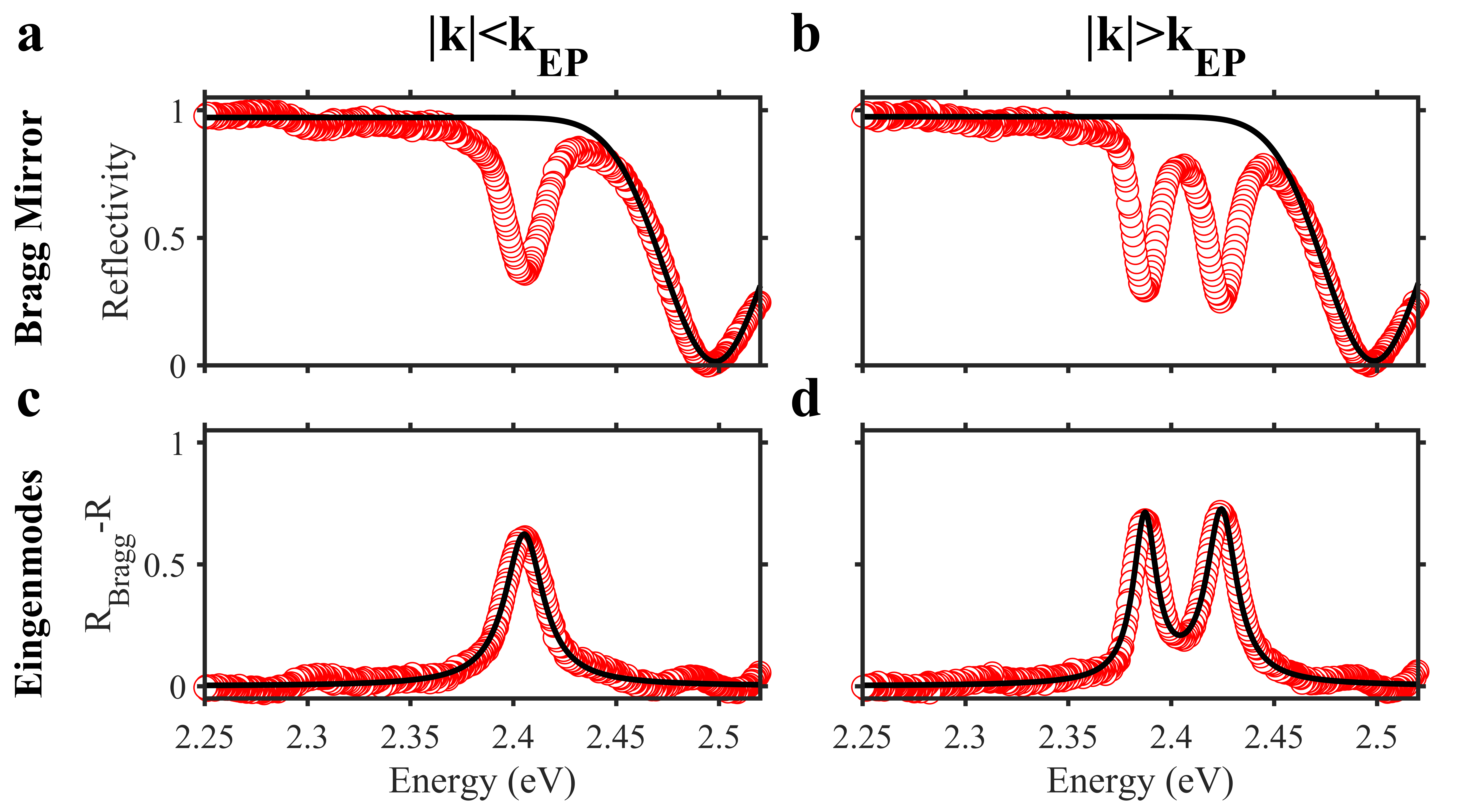}
  \caption{\textbf{Fitting of the vertical slices of the reflectivity map in figure 2 b.} \textbf{a} and \textbf{c} fit of the vertical slice at $|k|=0$ $\mu m^{-1}<k_{EP}$ (\textbf{a} extraction of the Bragg mirror reflectivity, \textbf{c} fit of the eigenmodes with one Lorentzian function). \textbf{b} and \textbf{d} fit of the vertical slice at $|k|=1.5$ $\mu m^{-1}>k_{EP}$ (\textbf{b} extraction of the Bragg mirror reflectivity, \textbf{d} fit of the eigenmodes with two Lorentzian functions). }
  \label{FigureD}
\end{figure*}

In order to collect the eigenvalues real parts (energies) and imaginary parts (linewidths) from the reflectivity map exhibiting exceptional points in figure 2 b, vertical slices were taken at given wave-vectors (see figures \ref{FigureD}). \\

The first step consists in extracting the Bragg mirror reflectivity by fitting the vertical slices with Gaussian functions (see figures \ref{FigureD} \textbf{a} and \textbf{b}). Then, the signal $R_{Bragg}-R$ are fitted with one or two Lorentz functions (see figures \ref{FigureD} \textbf{c} and \textbf{d}), with $R_{Bragg}$ the extracted Bragg mirror reflectivity, and $R$ the total reflectivity. Between the exceptional wavevector, i.e. for $|k|<k_{EP}$, the signals $R_{Bragg}-R$ were fitted with only one Lorentzian function as the two eigenmodes resonances merge into one due to the proximity of the eigenmodes energies (see figure \ref{FigureD} \textbf{c}). For wavevectors larger than the exceptional wavevector, i.e. for $|k|>k_{EP}$, the signals $R_{Bragg}-R$ were fitted with two Lorentzian functions (see figure \ref{FigureD} \textbf{d}). \\

We note that the eigenmodes resonances can be described by Lorentzian functions within the stop-band of the Bragg mirror. Indeed, the resonances are symmetric and the baseline specular reflectivity of the sample is close to one only within the Bragg mirror stop-band. \\

\cleardoublepage

\section{Collection of the eigenvalues from the reflectivity map with positive detuning} 

\begin{figure*}[!ht]
\centering
\includegraphics[width=1\linewidth]{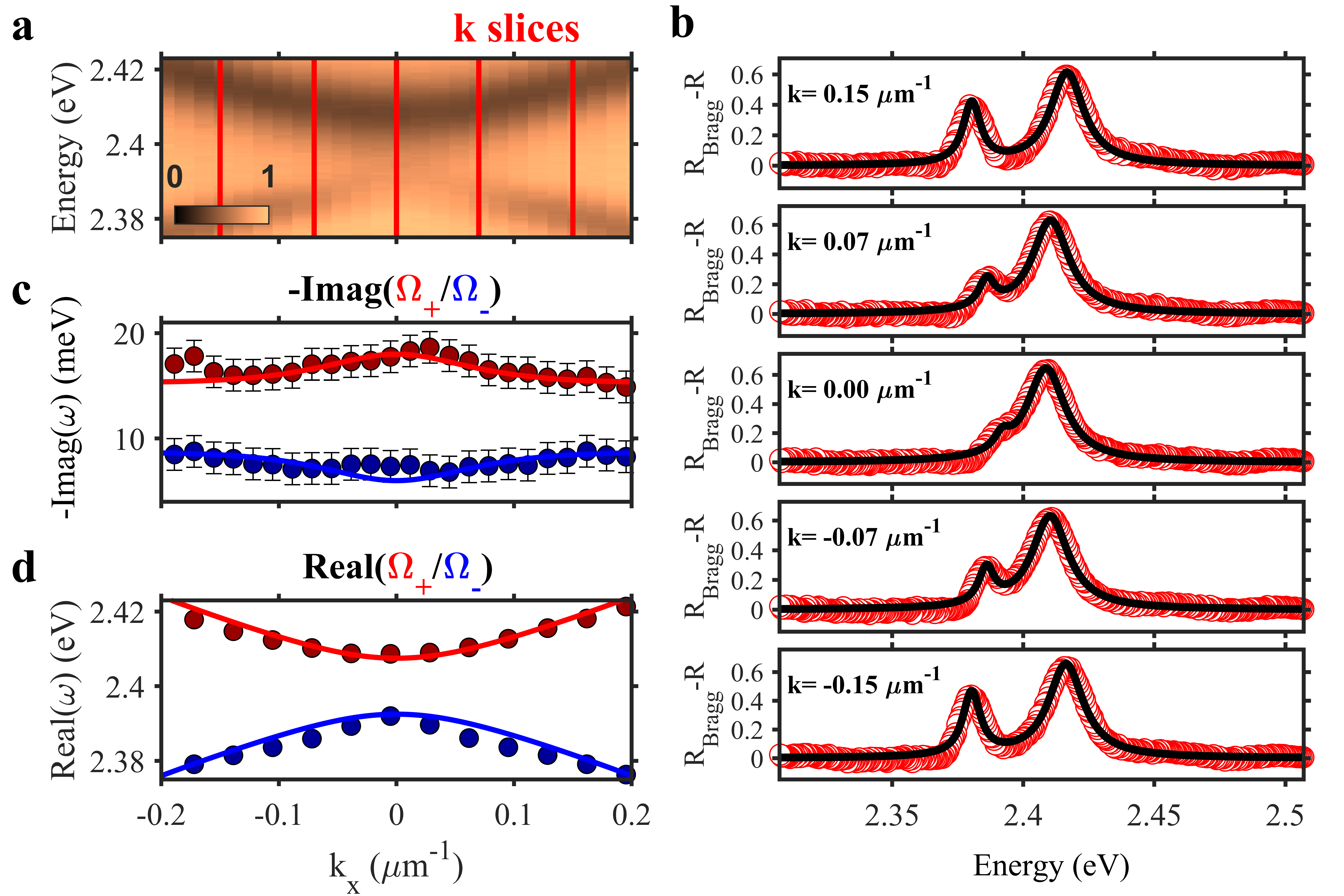}
  \caption{\textbf{Collection of the eigenvalues from the reflectivity map with negative detuning} \textbf{a} Angle-resolved reflectivity (ARR) pseudo-colour map, presented in figure 2 a with $\Delta<\Delta_{EP}$. The red vertical lines correspond to the vertical slices at given wave-vectors, $k$, shown in b. \textbf{b} Fitting of the eigenmodes resonances of five vertical slices taken at different wave-vectors: $k=$ -0.15, -0.07, 0.00, 0.07, and 0.15 $\mu m^{-1}$. \textbf{c} Real parts and imaginary parts of the experimental eigenvalues (blue and red dots) obtained from the reflectivity maps in figure 2 a compared with the model (blue and red lines) given in equation 2. The parameters used are $\Delta= -21 \: meV$, $\omega_0=2.4 \: eV$, $v=100 \: meV.\mu m^{-1}$, $\delta= -300 \: meV .\mu m^{-2}$, $\gamma_b=6 \: meV$, $\gamma_{nr}=3 \: meV$.}
  \label{FigureE}
\end{figure*}

 A similar study on the eigenvalues real and imaginary parts done for the reflectivity map exhibiting exceptional points was performed on the reflectivity map with a negative energy gap $\Delta$ shown in figure 2 \textbf{a}. The reflectivity map is reproduced in figure \ref{FigureE} \textbf{a} and the fitting of the vertical slices is shown in figure \ref{FigureE} \textbf{b}.  \\
 
 The obtained experimental eigenvalues real and imaginary parts are shown as blue and red dots in figure \ref{FigureE} \textbf{c} and \textbf{d}. The experimental results were fitted with the model given in equation 2 using the fitting parameters $\omega_0$, $\Delta$, $\delta$, $\gamma_b$, and $\gamma_{nr}$, while the group velocity, $v$ was directly measured to be $100 \: meV.\mu m^{-1}$ from the slope of the modes. A good agreement between the experimental results and the model is met with $\Delta= -21 \: meV$, $\omega_0=2.4 \: eV$, $\delta= -300 \: meV .\mu m^{-2}$, $\gamma_b=6 \: meV$, $\gamma_{nr}=3 \: meV$. Except for the detuning $\Delta$, the obtained fitting parameters are similar to the ones found for the reflectivity map exhibiting exceptional points. The good agreement between the experimental data and the model, and the consistency of the obtained fitting parameters, confirm the direct observation of the exceptional points. \\

\newpage

\section{Encircling the exceptional points}

\begin{figure*}[!ht]
\centering
\includegraphics[width=\linewidth]{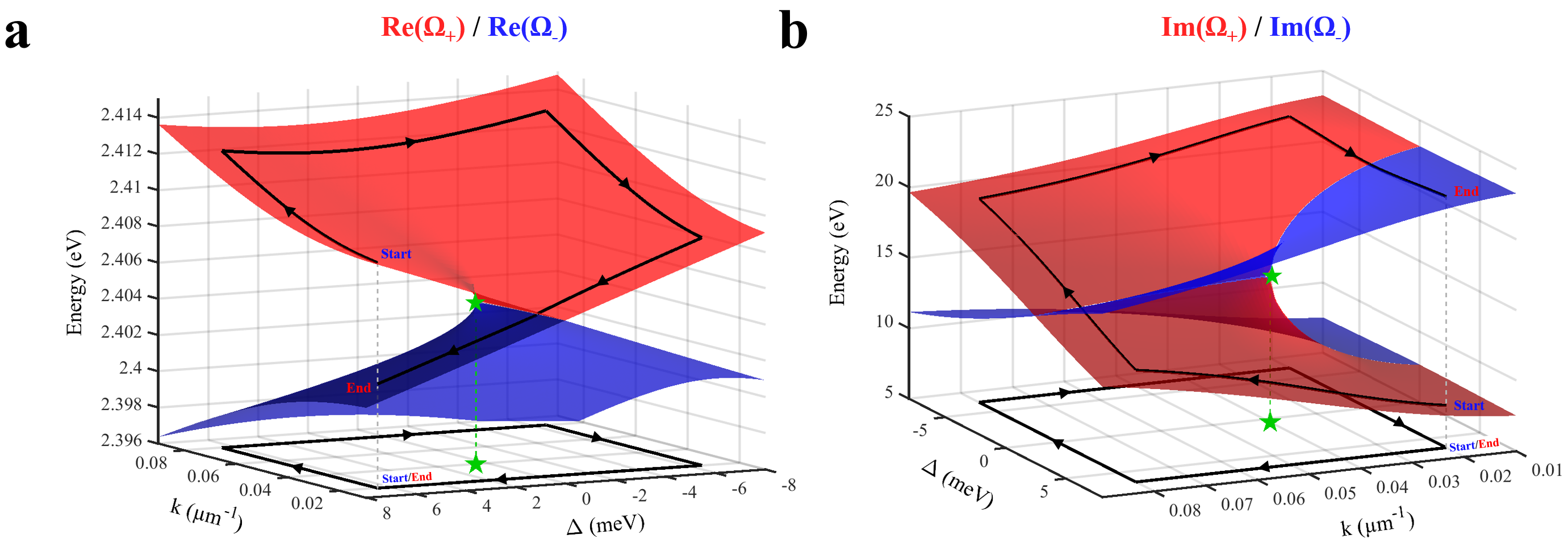}
  \caption{\textbf{Encircling an isolated exceptional point - Analytical model}. \textbf{a} Real part and \textbf{b} imaginary part of the analytical eigenvalues in the vicinity of an isolated exceptional point. Encircling the exceptional point in the parameters space results in switching the eigenvalues real and imaginary parts from the start point (in blue) and the end point (in red) after a one-trip turn.}
  \label{FigureI}
\end{figure*}

\begin{figure*}[!ht]
\centering
\includegraphics[width=\linewidth]{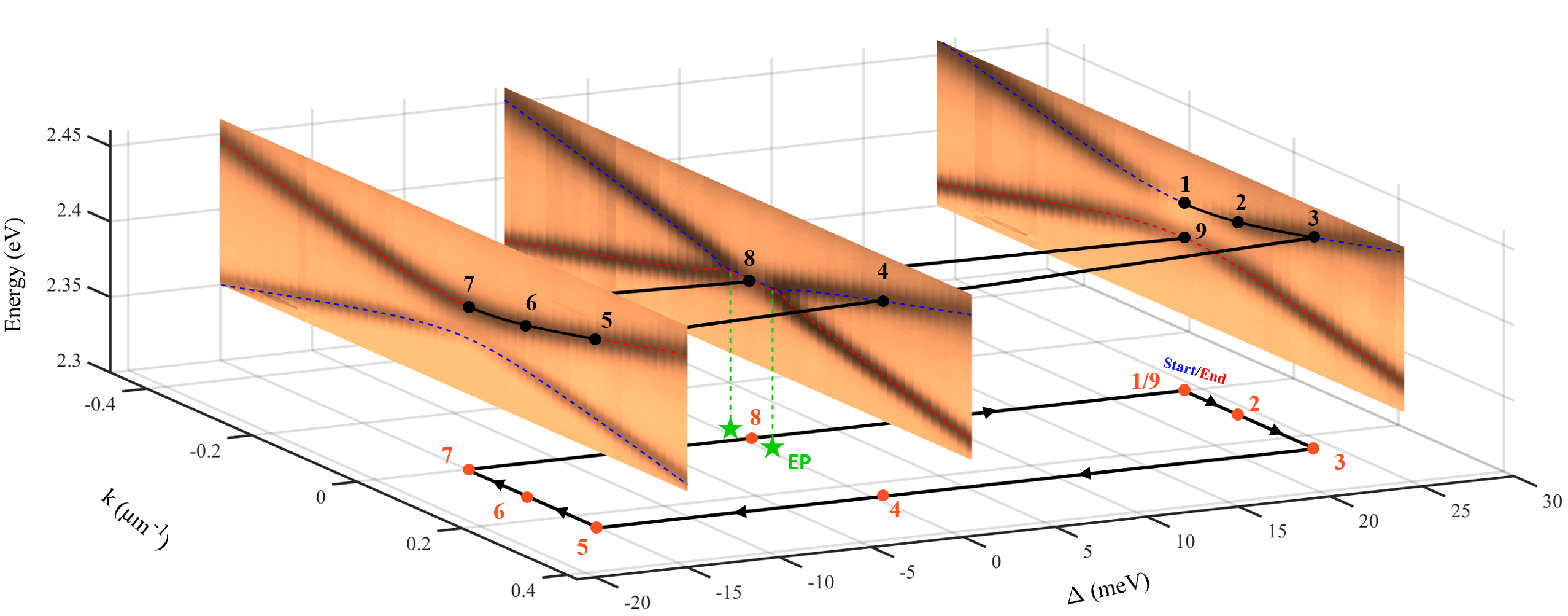}
  \caption{\textbf{Encircling an isolated exceptional point - Experiment}. Experimental angle-resolved reflectivity maps from figure 2 displayed as a function of the wavevector $k$ and detuning $\Delta$. An encirclement of an isolated exceptional point composed of 9 points is proposed and results in switching the eigenstates from the dark mode (1) to the bright mode (9).}
  \label{FigureJ}
\end{figure*}

\begin{figure*}[!ht]
\centering
\includegraphics[width=\linewidth]{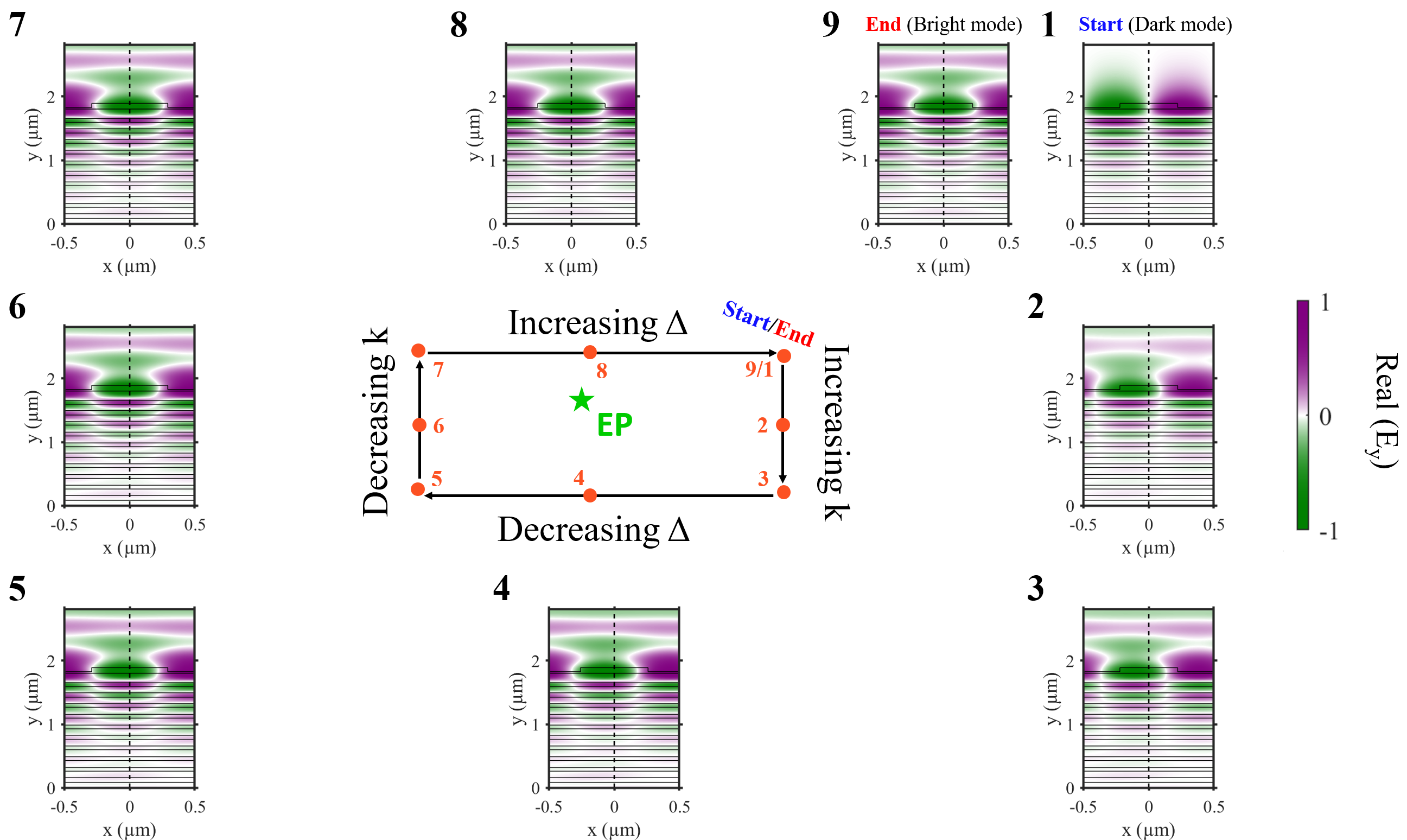}
  \caption{\textbf{Encircling an isolated exceptional point - Simulation}. Electric field distribution simulation performed with Comsol of the 9 points in the ($k$,$\Delta$) parameter space shown in figure \ref{FigureJ}. After a one-turn trip the anti-symmetric dark mode (1) is switched into the symmetric bright mode (9).}
  \label{FigureK}
\end{figure*}

\newpage

\section{Treatment of the photoluminescence data and calculation of the experimental LDOS}

\begin{figure*}[!ht]
\centering
\includegraphics[width=\linewidth]{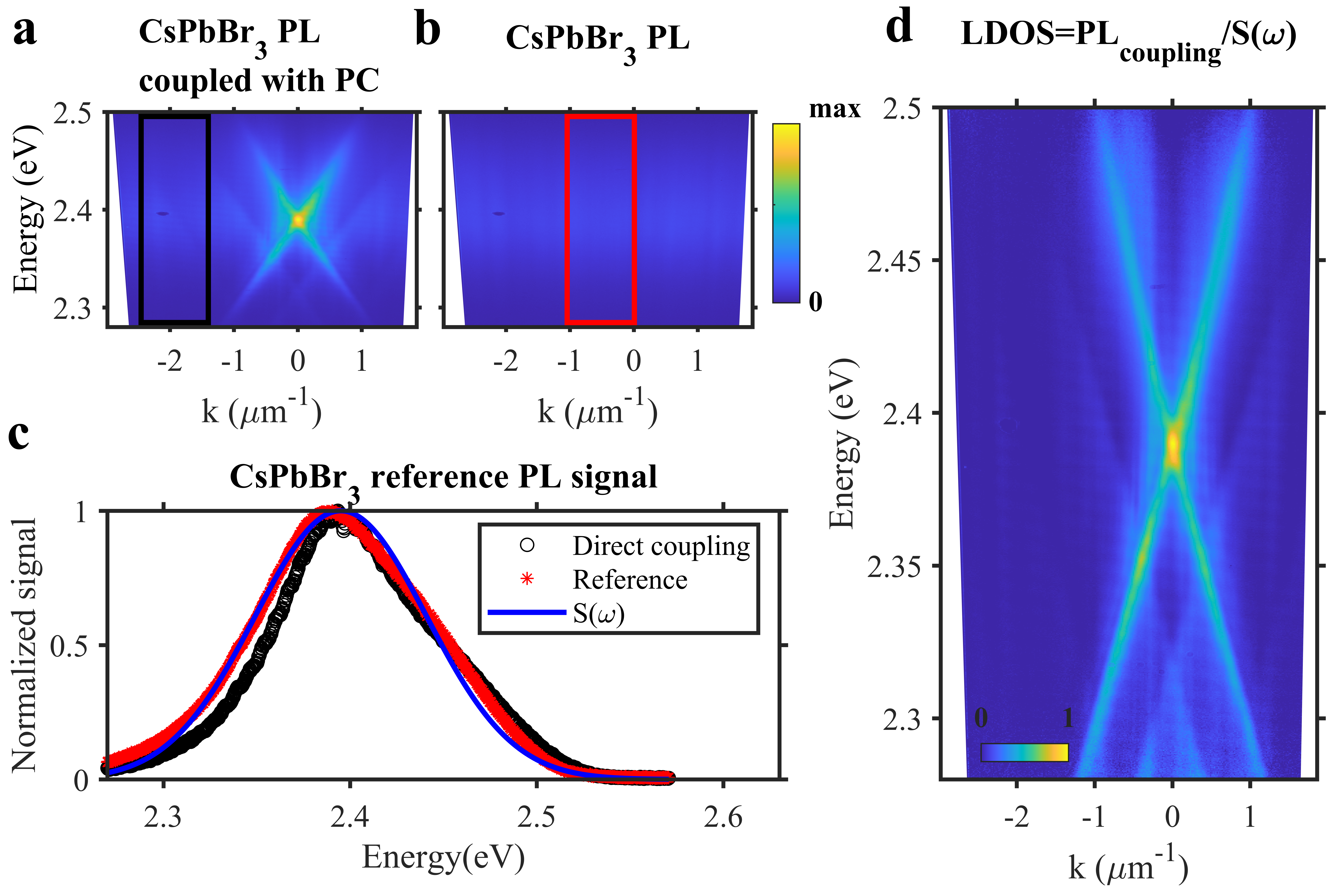}
  \caption{\textbf{Treatment of the photoluminescence data and calculation of the experimental LDOS} \textbf{a} Angle-resolved photoluminescence of the active sample. The photoluminescence signal corresponds to the perovskite \ch{CsPbBr_3} photoluminescence coupled with the modes of the grating, or 1D photonic crystal (PC). \textbf{b} Angle-resolved photoluminescence of the perovskite \ch{CsPbBr_3} uncoupled with the 1D photonic crystal (PC). The photoluminescence was measured in a region of the sample in which the perovskite was deposited away from the sample grating. \textbf{c} Integrated photoluminescence spectra of the perovskite \ch{CsPbBr_3} from the black region in a and the red region in b. The two spectra are fitted by a Gaussian function $S(\omega)$. \textbf{d} Experimental LDOS obtained extracted by dividing the normalized photoluminescence signal coupled to the photonic crystal modes, $I_{PL}^{Coupling}$, by the \ch{CsPbBr_3} normalized photoluminescence spectrum, $S(\omega)$ (see Eq. \ref{eqS1}). The signal coupled to the modes, $I_{PL}^{Coupling}(\omega,k)$, is obtained by subtracting the direct coupling signal, $I_{PL}^{DC}(\omega)$, to the total signal, $I_{PL}(\omega,k)$. }
  \label{FigureG}
\end{figure*}

To experimentally probe the Localized Density Of State (LDOS) at exceptional points, angle-resolved photoluminescence measurements (see section 4 of the supplementary) were performed on the active sample in a region corresponding to the grating in which the perovksite \ch{CsPbBr_3} is deposited on the structure (see figure \ref{FigureG} \textbf{a}), and a second region where the perovskite is only deposited in the sample substrate (see figure \ref{FigureG} \textbf{b}). \\

%with a pulsed pico-second laser (\textcolor{red}{xx - Hai Son Nguyen} KhZ, \textcolor{red}{xx - Hai Son Nguyen} ps rep. rate)

The direct coupling signal, $I_{PL}^{DC}(\omega)$ (photoluminescence signal which couples directly to the radiative continuum instead of the photonic crystal modes) is collected by integrating the photoluminescence signal delimited by the black box in figure \ref{FigureG} \textbf{a}. The reference photoluminescence spectrum of the perovskite \ch{CsPbBr_3} is obtained by integrating the photoluminescence signal delimited by the red box in figure \ref{FigureG} \textbf{b}. Figure \ref{FigureG} \textbf{c} shows the direct coupling (black) and reference (red) spectra fitted by a Gaussian function $S(\omega)$ (blue). \\

The LDOS is then extracted by dividing the normalized photoluminescence signal coupled to the photonic crystal modes, $I_{PL}^{Coupling}$, by the \ch{CsPbBr_3} normalized photoluminescence spectrum, $S(\omega)$:

\begin{equation}
\label{eqS1}
\centering
\begin{split}
&LDOS(\omega,k)=\frac{I_{PL}^{Coupling}(\omega,k)}{S(\omega)} \:, \\
\text{with} \quad
&I_{PL}^{Coupling}(\omega,k)=I_{PL}(\omega,k)-I_{PL}^{DC}(\omega) \:,
\end{split}
\end{equation} \\

\noindent where the signal coupled to the modes, $I_{PL}^{Coupling}(\omega,k)$, is obtained by subtracting the direct coupling signal, $I_{PL}^{DC}(\omega)$, to the total signal, $I_{PL}(\omega,k)$. \\

\newpage

\newpage 
\newpage

\section{Effect of $\gamma_{nr}$ on the LDOS}

\begin{figure*}[!ht]
\centering
\includegraphics[width=0.84\linewidth]{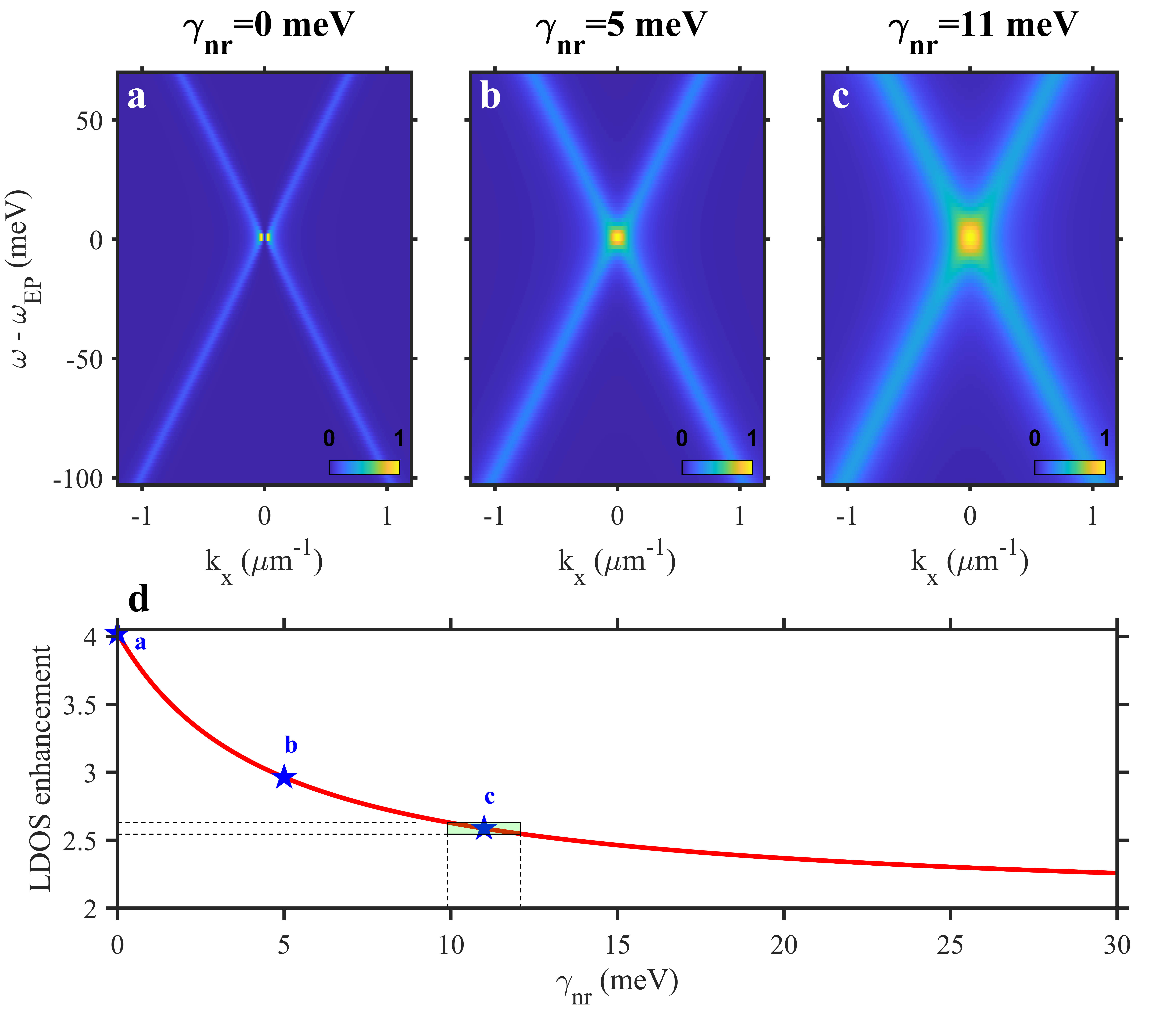}
 \caption{\textbf{Effect of $\gamma_{nr}$ on the LDOS.} \textbf{a} to \textbf{c} Theoretical LDOS maps obtained by using the model on the LDOS at EPs in \cite{Pick2017} and our analytical model for three different values of non-radiative losses $\gamma_{nr}$: (a) 0, (b) 5, and (c) 11 meV. Except for $\gamma_{nr}$ in (a) and (b), the used parameters are the same as in figure 3. These LDOS maps show the broadening in energy and wavevector of the LDOS peak at EPs for increasing non-radiative losses $\gamma_{nr}$. For non zero radiative losses $\gamma_{nr}$, the two LDOS peaks at EPs merge into one centered at $E=E_{EP}$ and $k=0 \mu m^{-1}$. \textbf{d} LDOS enhancement as a function of the non-radiative losses $\gamma_{nr}$. The LDOS enhancement is 4 at EPs without non-radiative losses ($\gamma_{nr}=0$ meV in a) and is reduced in presence of non-radiative losses $\gamma_{nr}$. The green interval of $\pm 10\%$ of $\gamma_{nr}$=11\,meV, corresponding to a variation of $\pm 10\%$ of the layer thicknesses, is indicated by the green shaded box. In this interval the enhancement factor varies from 2.55 to 2.63. }
  \label{Figure_LDOS_gamma_nr}
\end{figure*}

As discussed in the subsection \textit{"Theory of LDOS at Exceptional Points"} in the supplemental material, when there is no radiative losses, the passive enhancement factor is always four-fold for any slab thickness. However, this enhancement factor is reduced in presence of non-radiative losses in the system. Such an effect is shown in Fig.\ref{Figure_LDOS_gamma_nr} showing the theoretical LDOS maps obtained by using the model on the LDOS at EPs in \cite{Pick2017} and our analytical model for different values of non-radiative losses $\gamma_{nr}$. Except for $\gamma_{nr}$, the used parameters are the same as in figure 3. These LDOS maps show the broadening in energy and wavevector of the LDOS peak at EPs for increasing non-radiative losses $\gamma_{nr}$. \\ 

The enhancement factor 2.56 of our system corresponds to a total nonradiative loss-rate $\gamma_{nr}=$ 11\,meV (see figures \ref{Figure_LDOS_gamma_nr} \textbf{c} and \textbf{d}). This nonradiative loss is due to residual absorption in the HSQ grating ($\gamma_{nr}^{HSQ}\approx$ 3.8 \,meV extracted from passive measurement) and in the CsPbBr3 nanocrystals. One may expect that a variation of $\pm 10\%$ of the whole thickness (HSQ + nanocristals)  would lead to a $\pm 10\%$ variation of $\gamma_{nr}$, leading to $\gamma_{nr}$ varying from 9.9\,meV to  12.1\,meV. Locating this interval in Fig. \ref{Figure_LDOS_gamma_nr} \textbf{d} (see green shaded box), we expect an enhancement factor in the range of 2.55 to 2.63.

\newpage
\newpage

%\bibliography{mybibsupp}
%\bibliographystyle{ieeetr}

\bibliography{biblio}
\end{document}